\documentclass[authoryear,a4paper,review]{elsarticle} 
\pdfoutput=1 

\journal{Journal of Economic Dynamics and Control}

\usepackage{amsmath,amssymb,graphicx,hyperref,setspace} 

\usepackage{natbib}
\usepackage[left=1in,right=1in,top=1.4in,bottom=1.4in]{geometry}
\hypersetup{
	pdftitle={The equivalent CEV volatility of the SABR model},
	pdfauthor={Jaehyuk Choi, Lixin Wu},
	pdfkeywords={Stochastic volatility, SABR model, CEV model, implied volatility},
	colorlinks=true,
	linkcolor=red,
	citecolor=blue,
	urlcolor=blue,
	bookmarksnumbered=true,
	pdfstartview=
}

\date{May 5, 2021}

\newcommand{\distequal}{\,{\buildrel d \over =}\,}
\newcommand{\rhoc}{\rho_\ast}
\newcommand{\betac}{\beta_\ast}
\newcommand{\betacpow}[1]{\beta_\ast^{#1}}
\newcommand{\vov}{\nu}

\newcommand{\norm}{\textsc{n}}
\newcommand{\bs}{\textsc{bs}}
\newcommand{\cev}{\textsc{cev}}
\newcommand{\ncx}{{\chi^2}}
\newcommand{\sabr}{\textsc{sabr}}

\newcommand{\atan}{\mathrm{atan}}
\newcommand{\asinh}{\mathrm{asinh}}
\newcommand{\acosh}{\mathrm{acosh}}

\newcommand{\qtext}[2][\quad]{#1\text{#2}#1}

\newtheorem{thm}{Theorem}
\newproof{pf}{Proof}
\newproof{pfcor}{Proof of the $\rho=0$ case of Corollary~\ref{cor}}
\newtheorem{cor}{Corollary}[thm]
\newtheorem{rem}{Remark}

\begin{document}
\begin{frontmatter}

\title{The equivalent constant-elasticity-of-variance (CEV) volatility\\ of the stochastic-alpha-beta-rho (SABR) model}

\author[phbs]{Jaehyuk Choi\corref{corrauthor}}
\ead{jaehyuk@phbs.pku.edu.cn}

\author[hkust]{Lixin Wu}
\ead{malwu@ust.hk}


\cortext[corrauthor]{Corresponding author \textit{Tel:} +86-755-2603-0568, \textit{Address:} Rm 755, Peking University HSBC Business School, University Town, Nanshan, Shenzhen 518055, China}

\address[phbs]{Peking University HSBC Business School,\\ University Town, Nanshan, Shenzhen 518055, China}
\address[hkust]{Department of Mathematics, The Hong Kong University of Science and Technology,\\ Clear Water Bay, Kowloon, Hong Kong, China}

\begin{abstract}
This study presents new analytic approximations of the stochastic-alpha-beta-rho (SABR) model. Unlike existing studies that focus on the equivalent Black--Scholes (BS) volatility, we instead derive the equivalent constant-elasticity-of-variance (CEV) volatility. Our approach effectively reduces the approximation error in a way similar to the control variate method because the CEV model is the zero vol-of-vol limit of the SABR model. Moreover, the CEV volatility approximation yields a finite value at a zero strike and thus conveniently leads to a small-time asymptotics for the mass at zero. The numerical results compare favorably with the BS volatility approximations in terms of the approximation accuracy, small-strike volatility asymptotics, and no-arbitrage region.
\end{abstract}
\end{frontmatter}

\section{Introduction} \noindent
The stochastic-alpha-beta-rho (SABR) model proposed by \citet{hagan2002sabr} is one of the most popular stochastic volatility models adopted in the financial industry. Its commercial success is owing to a few factors. The model is intuitive and parsimonious. It provides a flexible choice of \textit{backbone}, the trace of the at-the-money volatility against the spot price. Most importantly, \citet{hagan2002sabr} provide an analytic approximation of implied Black--Scholes (BS) volatility in closed form (hereafter, the HKLW formula), from which traders can readily convert to the option price using the BS formula.

The HKLW formula is an asymptotic expansion valid for small time to maturities (up to the first order in time) and near-the-money strike prices. Several authors have attempted to improve the HKLW formula. Based on the results of \citet{berestycki2004impvol}, \citet{obloj2007fine} corrects the leading order term of the HKLW formula. \citet{henry2005general} derives the same leading order term from the heat kernel under hyperbolic geometry. \citet{paulot2015asym} further provides a second-order approximation which is accurate in a wider region of strike prices at the cost of numerical integrations for the second-order term. Further, \citet{lorig2017lsv} obtain implied BS volatility up to the third order in time, which unfortunately is valid only near the money. 
A more accurate solution of the SABR model, however, requires large-scale numerical methods such as the finite difference method~\citep{hagan2014arbfree,park2014sabr,vonsydow2019benchop}, continuous time Markov chain~\citep{cui2018ctmc-sabr}, multidimensional numerical integration~\citep{henry2005general,islah2009sabr-lmm,antonov2013sabr,korn2013exact}, or Monte-Carlo simulation~\citep{chen2012low,cai2017sabr,choi2019nsvh}.

By nature, analytic approximation methods suffer two important drawbacks when some parameters or strike price go beyond the \textit{comfort zones} of the concerned methods: non-negligible price error from the true value and the occurrence of arbitrage opportunity. Nevertheless, these methods are still attractive to practitioners because they are fast and robust. Note that practitioners need to compute the prices and Greeks of thousands of European options (or swaptions) frequently during trading hours. The calibration of the model parameters to the observed volatility smile also requires fast option evaluation because the parameters must be found using iterative methods. The numerical methods mentioned above are computationally intensive and not fast enough to use for those purposes.

Fortunately, the errors in analytic approximations are not a significant issue for those who use the SABR model primarily to price and manage the risk of European options. Specifically, the model parameters should first be calibrated to the market prices of the options at several liquid strike prices near the money. Then, the calibrated model is used to price the options at other strike prices. In this sense, the SABR model serves as a tool to \textit{interpolate} (as well as \textit{extrapolate}) the volatility smile, meaning that the accuracy of the price formula is less a concern.

Arbitrage under the analytic approximations occurs because an absorbing boundary condition is actually not imposed at the origin, as the small-time asymptotics of the transition density does not \textit{feel} the boundary. The SABR process has a non-zero probability of hitting zero for $0<\beta<1$, and an absorbing boundary condition should be explicitly imposed at the origin for $0<\beta<1/2$ for the price process to be a martingale and arbitrage-free. For this reason, the analytic approximations exhibit arbitrage opportunity in the low-strike region. The arbitrage outbreak is still an important concern to options market makers; savvy hedge funds can exploit them by purchasing a butterfly of options with a negative premium. To avoid such trades, market makers carefully keep track of the lower bound of the arbitrage-free region (and often patch a different arbitrage-free model below the bound). Therefore, the degree of arbitrage should be yet another performance measure for testing newly proposed analytic approximations, as in \citet{obloj2007fine}, as much as the approximation error.

We contribute to the SABR model literature by proposing new analytic approximations that are more accurate and have a wider arbitrage-free strike region than existing studies. We derive the equivalent volatility under the constant-elasticity-of-variance (CEV) model, from which the option price is computed with the analytic CEV option price formula~\citep{schroder1989comp}. We provide two formulas for the equivalent CEV volatility as spin-offs from existing studies. The first one (Theorem~\ref{th:cev1}) is obtained by following the approximation method of \citet{hagan2002sabr}. The second one (Theorem~\ref{th:cev2}) is simplified from \citet{paulot2015asym}'s original CEV volatility formula.

Our CEV-based approach is motivated by the simple intuition that the SABR model should converge to the CEV model when the volatility of volatility (vol-of-vol) approaches zero. Such motivation for using the CEV model is not novel in the SABR model literature. \citet{yang2017cev-sabr} show that the CEV option price (with the CEV volatility being the initial SABR volatility) is a good approximation in certain parameter ranges and is, naturally, arbitrage-free. The practical use of the result, however, is limited because the parameters related to the volatility process (i.e., vol-of-vol and correlation) are ignored in the approximation and only one degree of freedom is left to fit the volatility smile. Our work extends \citet{yang2017cev-sabr} as our CEV approximations have full dependency on the SABR parameters. \citet[\S 3.6, 4.5]{paulot2015asym} also discusses the equivalent CEV volatility as an alternative to BS volatility and outlines the derivation. His emphasis, however, is placed on BS volatility and the CEV volatility approximation is not tested numerically. Further, there is no discussion about the implications such as the mass at zero. In short, the advantage of the CEV approach has not thus far been explored. We fill this research gap by advocating the use of CEV volatility for the SABR model.

The numerical results show that our CEV-based approximations are more accurate than the corresponding BS-based methods from which they stem. In particular, the presented CEV approaches are more accurate when the initial volatility is large. This finding complements \citet{paulot2015asym}'s refinement, which makes the approximation more accurate for large vol-of-vol. Having both advantages, the second CEV approximation based on \citet{paulot2015asym} performs the best among all the methods over wide parameter ranges. In the numerical test for comparing the degree of arbitrage, the second CEV approximation also performs the best among all the methods in that negative implied probability density starts to appear at the lowest strike price (see Section~\ref{ssec:num-smallk}).

Surprisingly, the projection of the SABR model to the CEV model has the effect of imposing an absorbing boundary condition at the origin because the CEV price formula assumes the same boundary condition. Our CEV approximations offer finite CEV volatility at a zero strike, making them capable of implying the probability of hitting the origin. The mass-at-zero approximation in a closed-form formula (Theorem~\ref{th:mass}) shows excellent agreement with the numerical results in small time and is consistent with the exponentially vanishing asymptotics in the limit~\citep{chen2019feeling}. To the best of our knowledge, our approximation is the first closed-form approximation that works for all $\rho$, as the existing estimation methods either work on the uncorrelated case only~\citep{gulisashvili2018mass} or require numerical integration for the correlated case~\citep{yang2018survival}. Even if the mass at zero from our CEV approximations becomes less accurate in large time or vov-of-vol, our CEV approximations remain internally consistent with the model-free smile shape determined by the (possibly incorrect) mass at zero~\citep{demarco2017shapes}. This explains why our CEV approach has a wider arbitrage-free region.

The remainder of this paper is organized as follows. Section~\ref{sec:model} reviews the SABR model and existing BS volatility approximations. Section~\ref{sec:cev} derives the equivalent CEV volatility and mass-at-zero approximation. Section~\ref{sec:num} presents the numerical results. Finally, Section~\ref{sec:conc} concludes.

\section{SABR model and analytic approximations} \label{sec:model} \noindent
In this section, we introduce the SABR model and review various analytic approximation methods for the equivalent BS volatility in the order of increasing accuracy. Rather than simply repeating existing results, we reorganize the formulas in an insightful way, which should lead to our new approach in Section~\ref{sec:cev}.

\subsection{Model and standardization} \noindent
The stochastic differential equations (SDE) for the SABR volatility model~\citep{hagan2002sabr} are
\begin{equation}
\frac{dF_t}{F_t^\beta} = \sigma_t\, dW_t, \quad
\frac{d \sigma_t}{\sigma_t} = \vov \, dZ_t, \qtext{and}
dW_t\,dZ_t = \rho\, dt,
\end{equation}
where $F_t$ and $\sigma_t$ are the processes for the forward price and volatility, respectively, $\vov$ is the vol-of-vol, $\beta$ is the elasticity parameter, and $W_t$ and $Z_t$ are the standard Brownian motions (BM) correlated by $\rho$. Let $T$ be the time-to-maturity of the option; then, the SABR model is fully specified by the parameter set: $\{F_0, \sigma_0, \beta, \rho, \vov, T\}$. To simplify the notations, we also denote 
$$ \betac = 1 - \beta \qtext{and} \rhoc = \sqrt{1-\rho^2}.
$$

In order for the process to have a unique solution, an explicit boundary condition has to be specified for $0<\beta<1/2$ and it has to be an absorbing boundary condition for the price process to be a martingale and arbitrage-free. The absorbing boundary condition holds naturally for $1/2\le \beta < 1$. Then, the SABR model has a probability mass at the origin for $0<\beta<1$ similar to the CEV model. For the avoidance of doubt, no boundary condition is necessary for $\beta=0$ (the normal SABR model) because the price can freely go negative.

We first standardize the SDE to not only simplify the notations but also help with the intuition and numerical implementation. In particular, we standardize the price, strike, and volatility by their typical scales:
$$f_t=\frac{F_t}{F_0} \quad(f_0=1), \quad k=\frac{K}{F_0},\qtext{and} \hat{\sigma}_t = \frac{\sigma_t}{\sigma_0} \quad (\hat{\sigma}_0=1).$$
The SDE with standardized variables become
$$ \frac{df_t}{f_t^\beta} = \alpha \,\hat{\sigma}_t d W_t, \quad \frac{d\hat{\sigma}_t}{\hat{\sigma}_t} = \vov \,d Z_t, \qtext{where} \alpha = \frac{\sigma_0}{F_0^{\betac}}.
$$
Here, $\alpha$ is the initial volatility of the standardized price $f_t$ around $f_0=1$.\footnote{This should not be confused with the notation of other studies~\citep{hagan2002sabr,obloj2007fine,paulot2015asym}, where $\alpha$ is used for the initial volatility, $\alpha = \sigma_0$.} Indeed, the formula for $\alpha$ is a quick way of converting CEV volatility $\sigma_0$ to BS volatility $\alpha$ from $\alpha F_0 = \sigma_0 F_0^\beta$. Since $f_0=1$, $\alpha$ also serves as an approximation of normal and CEV volatility. As it turns out, $\alpha$ is the \textit{0-th} order term of the equivalent volatility of the SABR model in all the base models we consider: normal, BS, and CEV. In the rest of this paper, we use standardized variables, $k$ and $\alpha$, in the volatility approximations. In particular, the equivalent volatility (and its error) is presented as a ratio to $\alpha$. If $\sigma(k)$ is the equivalent volatility in the standardized scale (i.e., strike price $k$ and $f_0=1$), volatility in the original scale, $\sigma'(K)$, can be converted using
\begin{equation} \label{eq:c_scale}
\sigma'(K) = F_0^{1-\beta'} \sigma(k),
\end{equation}
where $\beta'$ is the elasticity parameter of the base model of the equivalent volatility (e.g., $\beta'=1$ for the BS model, $\beta'=0$ for the normal model, and $0<\beta'<1$ for the CEV model). For BS volatility, $\sigma(k)$ and $\sigma'(K)$ are same.

We can further reduce the dimension of the parameters by introducing a time scale. 
There are two choices of time scale: $T$ or $1/\vov^2$. Accordingly, the parameter set is reduced to $\{\alpha\sqrt{T}, \vov\sqrt{T}, \beta, \rho\}$ and $\{\vov/\alpha, \beta, \rho, \vov^2 T\}$, respectively. While the two parameter sets are equivalent, the first set better suggests the patterns of the asymptotic expansion of $T$; for example, we expect each order of $T$ in a small-time expansion to be accompanied by $\alpha^2$, $\alpha\vov$, and $\vov^2$.

\subsection{HKLW formula of \citet{hagan2002sabr}} \label{ssec:hklw} \noindent
In the original paper, \citet{hagan2002sabr} derive the equivalent BS volatility using the singular perturbation in the limit of a small time-to-maturity and a near-the-money strike price. They first derive the equivalent normal volatility of the SABR model as~\citep[Eq.~(A.59)]{hagan2002sabr}
\begin{subequations} \label{eq:hagan-norm}
\begin{equation}
\frac{\sigma_\norm}{\alpha} = \frac{\betac(k-1)}{k^{\betac}-1}\, H(\zeta) \big(1 + h_\norm\, T\big)
\qtext{for} \zeta = \frac{\vov}{\alpha} \frac{k-1}{k^{\beta/2}},
\end{equation}
where
\begin{equation} \label{eq:hagan-norm-h}
h_\norm = \frac{\betacpow{2}-1}{24\,k^{\betac}} \alpha^2 + \frac{\rho\beta}{4\,k^{\betac/2}} \alpha\vov + \frac{2-3\rho^2}{24} \vov^2,
\end{equation}
\end{subequations}
and the function $H(z)$ is defined in a chain as follows\footnote{The function $x(z)$ can be equivalently defined as $x(z) = -\log\left( \frac{V(z)-z-\rho}{1-\rho} \right)$. While the expression in Eq.~\eqref{eq:Hxv} is used by \citet{paulot2015asym}, the alternative expression, in the form of $-x(-z)$, is used by \citet{hagan2002sabr,hagan2014arbfree} and \citet{obloj2007fine}.}:
\begin{equation} \label{eq:Hxv}
H(z) = \frac{z}{x(z)}, \quad x(z) = \log\left( \frac{V(z)+z+\rho}{1+\rho} \right), \qtext{and} V(z) =\sqrt{1+2\rho z+z^2}.
\end{equation}
We use the dummy variable $z$ here because the functions will also be used with arguments other than $\zeta$. Regarding the evaluation of $H(z)$, there are two special cases to comment. At $z=0$, $H(z)$ should be evaluated as 1 from Taylor's expansion of $1/H(z)$ near $z=0$:
\begin{equation} \label{eq:Hseries}
\frac1{H(z)} = \frac{x(z)}{z} = 1 - \frac{\rho}{2}\,z + \frac{3\rho^2-1}{6}\,z^2 - \frac{(5\rho^2 - 3)\rho}8\, z^3
+ \;\cdots.
\end{equation}
When $\rho=0$, it can be simplified to $H(z) = z/\asinh(z)$ because $x(z) = \log (z + \sqrt{1+z^2}) = \asinh(z)$. The two properties will often be used later.

To obtain the equivalent BS volatility, \citet{hagan2002sabr} also derive the equivalent normal volatility of the BS model with volatility $\sigma_\bs$ as a special case of the SABR model with  $\alpha=\sigma_\bs$, $\beta=1$, and $\vov= 0$~\citep[Eq.~(A.63)]{hagan2002sabr}:
\begin{equation} \label{eq:norm-bs}
\frac{\sigma_\norm}{\sigma_\bs} = \frac{k-1}{\log k} \left(1 - \frac{\sigma_\bs^2}{24} T\right),
\end{equation} 
where $\log k$ appears as the limit of $(k^{\betac}-1)/\betac$ as $\betac\rightarrow 0$.
Then, Eqs.~(\ref{eq:hagan-norm}) and (\ref{eq:norm-bs}) are equated to solve for $\sigma_\bs$ up to $O(T)$:
\begin{subequations} \label{eq:hagan-bs}
\begin{equation} \label{eq:hagan-bs-vol}
\frac{\sigma_\bs}{\alpha} = \frac{\betac\log k}{k^{\betac}-1} H(\zeta) (1 + h_\bs\, T),
\end{equation}
where
\begin{equation} \label{eq:hagan-bs-h}
h_\bs \approx h_\norm + \frac{\sigma_\bs^2}{24} \approx \frac{\betacpow{2}}{24\,k^{\betac}} \alpha^2 + \frac{\rho\beta}{4\,k^{\betac/2}} \alpha\vov + \frac{2-3\rho^2}{24} \vov^2. 
\end{equation}
\end{subequations}
Here, $\sigma_\bs$ in Eq.~(\ref{eq:hagan-bs-h}) is replaced by the leading order approximation of Eq.~(\ref{eq:hagan-bs-vol}) near $k=1$,
$$ \sigma_\bs \approx \frac{\betac\log k}{k^{\betac}-1} \alpha \approx \frac{\alpha}{k^{\betac/2}}.
$$
The approximation comes from the expansion~\citep[Eq.~(A.68b)]{hagan2002sabr} near $k=1$,
\begin{equation} \label{eq:sinh}
\frac{k^{\betac} - 1}{\betac} = \frac{k^{\betac/2}}{\betac/2}\sinh\left(\,\log k^{\betac/2}\right) = k^{\betac/2} \log k \left(1+ \frac{\betacpow{2}}{24}\log^2 k +\frac{\betac^4}{1920}\log^4 k + \cdots \right).
\end{equation}
Further applying this expansion to $\betac\log k/(k^{\betac}-1)$ and $\zeta$, we finally arrive at the well-known HKLW formula~\citep[Eq.~(2.17)]{hagan2002sabr}:
\begin{equation} \label{eq:hklw}
\frac{\sigma_\bs}{\alpha} = \frac{H(\zeta')}{k^{\betac/2}} \, \frac{1 + \left(\frac{\betacpow{2}}{24\,k^{\betac}} \alpha^2 + \frac{\rho\beta}{4\,k^{\betac/2}} \alpha\vov + \frac{2-3\rho^2}{24} \vov^2\right) T }{1+ \frac{\betacpow{2}}{24}\log^2 k +\frac{\betac^4}{1920}\log^4 k}
\qtext{for} \zeta' = \frac{\vov}{\alpha} k^{\betac/2} \log k,
\end{equation}
where $H(\cdot)$ is defined in Eq.~\eqref{eq:Hxv}.

\subsection{Corrected leading order term of \citet{obloj2007fine}} \noindent
We first define $q$ and $z$ as 
\begin{equation} \label{eq:zq}
q = \int_1^k k^{-\beta}dk = 
\begin{cases}
\dfrac{k^{\betac}-1}{\betac} & \text{if}\;\; 0\le \beta < 1 \\
\;\;\log k & \text{if}\quad \beta=1
\end{cases}
\qtext{and}
z = \frac{\vov}{\alpha}q.
\end{equation}
Let us also separately define $q$ (and $z$) for the two special cases of $\beta=0$ and $1$:
\begin{equation} \label{eq:zq01}
q_\norm = k-1 \; \left(z_\norm = \frac{\vov}{\alpha}q_\norm\right) \qtext{and} q_\bs = \log k \; \left(z_\bs = \frac{\vov}{\alpha} q_\bs\right).
\end{equation} 
Therefore, $z\rightarrow 0$ in the zero vol-of-vol limit ($\vov\downarrow 0$) and $z=q=0$ at the money ($k=1$).

Based on the results of \citet{berestycki2004impvol}, \citet{obloj2007fine} corrects the leading order term $H(\zeta)$ of \citet{hagan2002sabr} by replacing $\zeta$ with $z$. After the correction, the equivalent normal and BS volatilities, Eqs.~\eqref{eq:hagan-norm} and \eqref{eq:hagan-bs}, respectively become
\begin{equation} \label{eq:obloj}
 \frac{\sigma_\norm}{\alpha} = \frac{q_\norm}{q} H(z) (1 + h_\norm\, T) \qtext{and} \frac{\sigma_\bs}{\alpha} = \frac{q_\bs}{q} H(z) (1 + h_\bs\, T), 
\end{equation}
where $h_\norm$ and $h_\bs$ are unchanged as defined in Eqs.~\eqref{eq:hagan-norm-h} and \eqref{eq:hagan-bs-h}, respectively. Both $q_\norm/q$ and $q_\bs/q$ are numerically evaluated as 1 at $k=1$.

We briefly explain the variable $z$ in this correction. With the scaled time, $s = t \,\vov^2$, the SABR SDE are equivalently stated as 
\begin{equation} \label{eq:SDE_std}
\frac{\vov}{\alpha} \frac{df_t}{f_t^\beta} = \hat{\sigma}_t d\hat{W}_s, \quad \frac{d\hat{\sigma}_t}{\hat{\sigma}_t} = d\hat{Z}_s \qtext{for} t = \frac{s}{\vov^2},
\end{equation}
where $\hat{W}_s$ is a standard BM rescaled from $W_t$ by $\hat{W}_s = (1/\vov) W_{s \vov^2}$ (same for $\hat{Z}_s$). The variable $z$ is the Lamperti transformation,
$$ z_t = \frac{\vov}{\alpha} \int_{f=1}^{f_t} \frac{df}{f^\beta} = \frac{\vov}{\alpha\betac}(f_t^{\betac}-1),
$$
evaluated with $f_t=k$. In fact, both $\zeta$ and $\zeta'$ are approximations to $z$ near $k=1$.

We deliberately denote the new state variable by the same $z$ as the dummy variable in Eq. (\ref{eq:Hxv}) because $z$ is the correct variable for the functions in Eq.~(\ref{eq:Hxv}). In the rest of this paper, we often omit the argument $z$ from the functions for the sake of conciseness (e.g., $H=H(z)$, $x = x(z)$, and $V=V(z)$), unless otherwise stated.

\subsection{Improved normal volatility approximation of \citet{hagan2014arbfree}} \noindent
While the HKLW formula, Eq~\eqref{eq:hklw}, is widely used as the final outcome of \citet{hagan2002sabr}, the normal volatility approximation, Eq.~\eqref{eq:hagan-norm}, is also popular among practitioners in fixed income trading, for which the SABR model was originally proposed. In fixed income trading, normal volatility is preferred to BS volatility for quoting and managing the risk of swaptions. Moreover, the normal volatility formula is considered to be more accurate because it avoids an extra step for approximating BS volatility.

In a follow-up paper, \citet[Eqs.~(14)--(16)]{hagan2014arbfree} present an equivalent normal volatility improved over Eq.~\eqref{eq:hagan-norm}:
\begin{subequations} \label{eq:hagan-new}
	\begin{equation}
	\frac{\sigma_\norm}{\alpha} = \frac{q_\norm}{q}\, H(z) \big(1 + h_\norm\, T\big),
	\end{equation}
	where $q$, $z$, and $q_\norm$ are from Eqs.~\eqref{eq:zq}--\eqref{eq:zq01}, $H(z)$ is from Eq.~\eqref{eq:Hxv}, and
	\begin{equation} \label{eq:hagan-new-h}
	h_\norm = \log \left(k^{\beta/2}\frac{q}{q_\norm}\right) \frac{\alpha^2}{q^2} + \frac{\rho}{4} \frac{k^\beta-1}{k-1} \alpha\vov + \frac{2-3\rho^2}{24} \vov^2.
	\end{equation}
\end{subequations}
This new approximation not only adopts the correction of \citet{obloj2007fine} in the leading order term but also further refines the first-order term, $h_\norm$. 
Rather than deriving the equivalent BS volatility, \citet{hagan2014arbfree} promote using the normal model~\citep{bachelier1900} with this normal volatility to obtain the option price. If needed, implied BS volatility can be quickly inverted from the price using an accurate approximation such as \citet{jackel2015let}. 

In particular, the normal SABR ($\beta=0$) has been a popular model choice to admit negative interest rates~\citep{antonov2015free} when the interest rate hovered near zero since the global financial crisis in 2008. From both Eqs.~\eqref{eq:hagan-norm} and \eqref{eq:hagan-new}, the normal volatility approximation of the normal SABR model is derived as
\begin{equation} \label{eq:norm}
\frac{\sigma_\norm}{\alpha} = H(z_\norm) \left(1 + \left(\frac{2-3\rho^2}{24} \vov^2\right) T\right).
\end{equation}
Here, several observations should be mentioned. For the normal SABR model, the equivalent normal volatility is a natural choice. Using the equivalent BS volatility would be non-sensical because it restricts the price to be non-negative, contradicting the motivation for choosing $\beta=0$. The leading order term is simply $H(z_\norm)$, meaning that the zero vol-of-vol limit is correct; $\sigma_\norm \rightarrow \alpha$ as $\vov \downarrow 0$. Moreover, Eq.~\eqref{eq:norm} does not have a divergence issue at $k=0$, unlike the general case of Eq.~\eqref{eq:hagan-new-h}. These observations are generalized to $0<\beta<1$ in our CEV approach in Section~\ref{sec:cev}.

\subsection{Refined first-order term of \citet{paulot2015asym}} \label{ssec:paulot} \noindent
Using the heat kernel expansion of hyperbolic geometry, \citet{paulot2015asym} derives an equivalent BS volatility up to $O(T^2)$. While \citet{henry2005general} also uses the heat kernel expansion to derive the correct leading order volatility, \citet{paulot2015asym} derives the $O(T)$ and $O(T^2)$ terms without approximating the dependency on the strike price in each time order, thereby making the equivalent volatility valid for a wider region of strike prices. Although the $O(T^2)$ term improves accuracy, we adopt the approximation only up to the order $O(T)$ because the $O(T^2)$ term involves a numerical integration which defeats the purpose as an analytic approximation. 

\citet{paulot2015asym} expresses the small-time asymptotics of option's time value under a stochastic volatility model in the general form:
\begin{equation} \label{eq:optval}
P(k) = \frac{\sigma \sqrt{T}}{\sqrt{2\pi}}\left(\frac{\sigma^2 T}{d^2}\right)\exp\left(-\frac{d^2}{2\sigma^2 T} + e + O(T)\right),
\end{equation} 
where $\sigma$ is the initial volatility of the model at $t=0$, $d$ is the geodesic distance between the initial ($f_0=1$) and final (i.e., $f_T=k$) points on the differential geometry characterized by the model, and $e$ is also to be determined from the model. Here, the time value is also understood as the out-of-the-money put option price, hence the notation $P(k)$, because we are concerned with the low-strike region ($k<1$). For example, the coefficients for the BS model with volatility $\sigma_\bs$ are given by~\citep[Eq.~(21)]{paulot2015asym}
\begin{equation} \label{eq:bs_ov}
\sigma = \sigma_\bs, \quad d_\bs = q_\bs \; (=\log k), \qtext{and} 
e_\bs = \log \sqrt{k}.
\end{equation}

The expansion of the option value under the SABR model in the short-time limit is given by~\citep[Eq.~(32)]{paulot2015asym} 
\begin{equation} \label{eq:sabr_ov}
\sigma = \alpha, \quad d_\sabr =\frac{q}{H} \left(= \frac{\alpha}{\vov}x\right), \qtext{and} 
e_\sabr = \frac{q^2}{\alpha^2} (A_2 + A_3) + \log\left(H k^{\beta/2} \right),
\end{equation}
where
\begin{equation} \label{eq:paulot-bs-h23}
A_2 = \begin{cases} \displaystyle
\frac{\beta\rho\vov^2}{\betac\rhoc}\; \frac{G(t_2)-G(t_1)}{z^2} & (0\le \beta<1) \vspace{0.3em}\\
\displaystyle
\frac{\rho\alpha\vov}{2\rhoc^2}\; \frac{V - 1-\rho z}{z^2} & (\beta=1)
\end{cases}
\qtext{and}
A_3 = \frac{\vov^2}{2 z^2} \log\left(\frac{V}{H^2} \right).
\end{equation}
Here, $t_1$, $t_2$, and $G(t)$ in $A_2$ are defined by
\begin{equation} \label{eq:t1t2} 
t_1 = \frac{V+z+\rho}{\rhoc}, \quad t_2 = \frac{1+\rho}{\rhoc},
\end{equation}
and
\begin{equation} \label{eq:G}
G(t) = \atan(t) + \begin{cases}
\frac{\eta}{2\sqrt{1 - \eta^2}}\, \log\left|
\frac{\rho + (\eta-\rhoc)t + \sqrt{1-\eta^2}}{\rho + (\eta-\rhoc)t - \sqrt{1-\eta^2}}
\right| & \quad (0\le \eta < 1) \\
\frac{1}{\rho + (1-\rhoc)t} & \quad (\eta = 1)\\
\frac{\eta}{\sqrt{\eta^2 - 1}}\, \atan\left(
\frac{\sqrt{\eta^2-1}}{\rho + (\eta-\rhoc)t}
\right) & \quad (\eta > 1)
\end{cases} \text{for}\quad \eta = \frac{\rhoc \vov k^{\betac}}{\betac \alpha \,V}.
\end{equation}
We have rearranged the original expressions in \citet{paulot2015asym} into $A_2$ and $A_3$ (and $A_1$ later) to handle the limit $k\rightarrow 1$ later. Moreover, we further simplify the expression of $A_2$, which was originally given in terms of several layers of definitions. \ref{apdx:simp} provides the details of the simplification. The rearrangement and simplification not only facilitate the numerical implementation but also help explain the formula, as we discuss below.

Next, the two option value expansions are equated to solve for the equivalent BS model. Assuming an expansion in $T$, $\sigma_\bs = \sigma_{\bs,0} (1+ h_\bs T)$, we obtain $\sigma_{\bs,0}$ and $h_\bs$ sequentially as follows:
\begin{gather*}
\frac{\sigma_{\bs,0}}{\alpha} = \frac{d_\bs}{d_\sabr} = \frac{q_\bs}{q} H, \\
\text{and}\quad h_\bs = \frac{\alpha^2}{d_\sabr^2}\left(e_\sabr - e_\bs + \log\left(\frac{\alpha}{\sigma_{\bs,0}}\right)\right) = H^2(A_1 + A_2 + A_3),
\end{gather*}
where
\begin{equation}
\label{eq:paulot-bs-h1}
A_1 = \log\left(\frac{q}{q_\bs}\, k^{-\betac/2} \right) \frac{\alpha^2}{q^2}.
\end{equation}
Finally, we arrive at the equivalent BS volatility formula of \citet{paulot2015asym}:
\begin{equation} \label{eq:paulot-bs}
\frac{\sigma_\bs}{\alpha} = \frac{q_\bs}{q} H(z)\, \big(1+ h_\bs T\big) \qtext{for} h_\bs = H(z)^2\left(A_1 + A_2 + A_3\right),
\end{equation}
where $H(z)$ is defined in Eq.~\eqref{eq:Hxv}; $q$, $z$, and $q_\bs$ are defined in Eqs.~\eqref{eq:zq}--\eqref{eq:zq01}; and $A_1$, $A_2$, and $A_3$ are defined in Eqs.~\eqref{eq:paulot-bs-h1} and \eqref{eq:paulot-bs-h23}.

It is worthwhile checking the circumstances in which the refinement of \citet{paulot2015asym} makes a difference to the approximations of \citet{hagan2002sabr} with \citet{obloj2007fine}'s correction. To begin with, the leading order term, $(q_\bs/q)H(z)$, is the same as that of \citet{obloj2007fine}. Although the expressions of $h_\bs$ in Eqs.~\eqref{eq:hagan-bs} and \eqref{eq:paulot-bs} look different, they have the same value at the money ($k=1$)\footnote{This is stated by \citet[\S 4.3]{paulot2015asym} without proof.}. Indeed, we have purposely introduced $A_1$, $A_2$, and $A_3$ to decompose $h_\bs$ in such a way that the three terms correspond to those in Eq.~\eqref{eq:hagan-bs-h}, respectively:
$$ A_1 \rightarrow \frac{\betacpow{2}}{24}\,\alpha^2, \quad A_2 \rightarrow \frac{\rho\beta}{4} \alpha\vov, \qtext{and} A_3 \rightarrow \frac{2-3\rho^2}{24} \vov^2 \qtext{as} k\rightarrow 1.
$$
The limits of $A_1$ and $A_3$ can be easily derived from the expansions, Eqs.~(\ref{eq:sinh}) and (\ref{eq:Hseries}), respectively. The limit of $A_2$ requires additional algebra, which is placed in \ref{apx:converge}. Therefore, \citet{paulot2015asym} is distinguished from \citet{hagan2002sabr} and \citet{obloj2007fine} only for out-of-the-money strike prices ($k\neq 1$). Since the dependency on $k$ is manifested through $z=(\vov/\alpha)\,q$, we also expect \citet{paulot2015asym}'s refinement to be more pronounced when the $\vov/\alpha$ ratio is large.

\subsection{Low-strike smile and mass at zero of the BS-based approximations} \label{ssec:mass-bs} \noindent
We comment on the low-strike behavior of the BS volatility approximations. In general, analytic approximation does not take into account the boundary condition at the origin because the small-time asymptotics of the transition density ignores the boundary condition. Therefore, asymptotic approximation should not be trusted near $k=0$.

As $k$ approaches zero, approximation quality deteriorates and eventually causes negative implied probability density, inducing arbitrage. The occurrence of arbitrage can also be seen through the fact that the equivalent BS volatilities we have reviewed do not conform to the model-free volatility bound of \citet{lee2004moment} for the small strike: $\sqrt{2|\log k|/T}$ as $k\downarrow 0$ for all $T$. The corrected leading order of the equivalent BS volatilities scale as $(q_\bs/q)H(z)\sim O(|\log k|)$ as $k\downarrow 0$ for $0<\beta<1$, and this breaches the upper bound of \citet{lee2004moment}.

Furthermore, the probability of hitting the origin is critical for understanding the volatility smile in the low-strike region. \citet[Eq.~(1.4)]{demarco2017shapes} states that in the presence of the mass at zero, the model-free BS volatility smile at a small strike is governed by the mass at zero, $M_T = \mathbb{P}(f_T=0)$:
\begin{equation} \label{eq:demarco}
\sigma_\bs = \frac{L}{\sqrt{T}} \left( 1 + \frac{q}{L} + \frac{q^2+2}{2L^2} + \frac{q}{2L^3} + \cdots \right) \qtext{where} L=\sqrt{2|\log k|\,} \text{ and } q = N^{-1}(M_T),
\end{equation} 
where $N^{-1}(\cdot)$ is the inverse cumulative distribution function (CDF) of the normal distribution. The first term in Eq.~\eqref{eq:demarco} corresponds to \citet{lee2004moment}'s upper bound.

Similar to the CEV model, the SABR model also has the probability mass at the origin for $0<\beta<1$. Therefore, the small-strike smile under the SABR model is subject to Eq.~\eqref{eq:demarco}. Several researches estimate the mass at zero under the SABR model and, eventually, obtain the small-strike smile by taking advantage of Eq.~\eqref{eq:demarco}. \citet{gulisashvili2018mass} derive the approximations for $M_T$ for the uncorrelated case ($\rho=0$) in small- and large-time limits. \citet{yang2018survival} derive $M_T$ for the correlated case in the small-time limit. The practical use of the formulas, however, is limited because Eq.~\eqref{eq:demarco} is valid for a small strike (diverges at $k=1$) and it is non-trivial to merge the small-strike smile into the approximations reviewed earlier that are valid near the money.

\section{CEV-based approximation} \label{sec:cev}
\subsection{CEV model} \noindent
Since we advocate the use of implied CEV volatility, we briefly review the CEV model. The standardized CEV model with volatility $\sigma_\cev$ is given by
$$ \frac{d f_t}{f_t^\beta} = \sigma_\cev\, dW_t \quad (f_0=1). $$
The standardized prices of the call and put options with strike price $k$ and time-to-maturity $T$ are respectively~\citep{schroder1989comp}
\begin{gather}
C_\cev(k) = \bar{F}_\ncx \left(\frac{k^{2\betac}}{\betacpow{2}\sigma_\cev^2 T}; \,2+\frac1{\betac},\frac{1}{\betacpow{2}\sigma_\cev^2 T}\right) - k \, F_\ncx \left(\frac{1}{\betacpow{2}\sigma_\cev^2 T};\, \frac1{\betac},\frac{k^{2\betac}}{\betacpow{2}\sigma_\cev^2 T}\right), \label{eq:cev_call}\\
P_\cev(k) = k \, \bar{F}_\ncx \left(\frac{1}{\betacpow{2}\sigma_\cev^2 T};\, \frac1{\betac},\frac{k^{2\betac}}{\betacpow{2}\sigma_\cev^2 T}\right) - F_\ncx \left(\frac{k^{2\betac}}{\betacpow{2}\sigma_\cev^2 T}; \,2+\frac1{\betac},\frac{1}{\betacpow{2}\sigma_\cev^2 T}\right),\label{eq:cev_put}
\end{gather}
where $F_\ncx(x;r, x_0)$ and $\bar{F}_\ncx(x;r,x_0)$ are respectively the CDF and complementary CDF of the non-central chi-squared distribution with degrees of freedom $r$ and non-centrality parameter $x_0$. The Greeks, namely, the sensitivity of the price with respect to the parameters, are also analytically available; see \citet{larguinho2013cev}. The functions related to the non-central chi-squared distribution are included in many standard numerical libraries. Several approximation methods~\citep{sankaran1963approx,fraser1998approx} are also available to speed up the evaluation; see \citet{larguinho2013cev} for further details. 

Under the CEV model, the absorbing boundary condition has to be imposed at the origin in order to make the price process a martingale and arbitrage-free. The option formulas above are indeed derived with the absorbing boundary condition. Eqs.~\eqref{eq:cev_call}--\eqref{eq:cev_put} imply $C_\cev(0) = 1$ and $P_\cev(0) = 0$, satisfying the put--call parity at $k=0$. The CDF of the price distribution is given by
$$ \mathbb{P}(f_T\le k) = \bar{F}_\ncx \left(\frac{1}{\betacpow{2}\sigma_\cev^2 T};\, \frac1{\betac},\frac{k^{2\betac}}{\betacpow{2}\sigma_\cev^2 T}\right).
$$
In particular, the mass at zero for $0<\beta<1$ is analytically available as
\begin{equation} \label{eq:cev_m0}
M_T = \mathbb{P}(f_T=0) = \bar{F}_\ncx\left(\frac{1}{\betacpow{2}\sigma_\cev^2 T};\, \frac1{\betac},0\right) = 
\bar{\Gamma}\left(\frac{1}{2\betacpow{2}\sigma_\cev^2 T};\,\frac{1}{2\betac} \right),
\end{equation}
where $\bar{\Gamma}(x;a)$ is the complementary CDF of the gamma distribution\footnote{This is equivalent to the upper incomplete gamma function normalized by the gamma function $\Gamma(a)$.} with the shape parameter $a>0$. The definition of $\bar{\Gamma}(x;a)$ and its asymptotic expansions for large $x$ \citep[(6.5.32)]{abramowitz} are respectively given by
\begin{align} 
\bar{\Gamma}(x;a) &= \frac{1}{\Gamma(a)}\int_x^\infty t^{a-1} e^{-t} dt \label{eq:gammasf}\\ 
&\sim \frac{x^{a-1}\,e^{-x}}{\Gamma(a)} \left(1 + \frac{a-1}{x} + \frac{(a-1)(a-2)}{x^2} + \cdots \right) \qtext{as} x\rightarrow \infty. \label{eq:gammasf_asymp}
\end{align}

For later use, we summarize the small-strike asymptotics of the put option price under the CEV model. If a price distribution in general is defined to be non-negative, the CDF and put option price, $P(k)$, satisfy
\begin{equation} \label{eq:bound}
M_T \le \frac{P(k)}{k} \le \mathbb{P}(f_T\le k),
\end{equation}
because the term in the middle is the price of the put spread struck at $0$ and $k$ with $P(0)=0$. Therefore, if $M_T>0$, $$ M_T \sim \frac{P(k)}{k} \qtext{or} P(k) \sim k\, M_T \qtext{as} k\downarrow 0. $$
In the CEV model context, for all $T$ and $0<\beta<1$, we have
\begin{equation} \label{eq:cev_lowk}
P_\cev(k) \sim k M_T =
k\, \bar{\Gamma}\left(\frac{1}{2\betacpow{2}\sigma_\cev^2 T};\,\frac{1}{2\betac} \right) \qtext{as} k \downarrow 0.
\end{equation}
It can also be directly shown from Eq.~\eqref{eq:cev_put} using the fact that the second term vanishes faster than $O(k)$:
\begin{equation} \label{eq:cev_2nd}
F_\ncx \left(\frac{k^{2\betac}}{\betacpow{2}\sigma_\cev^2 T}; \,2+\frac1{\betac},\frac{1}{\betacpow{2}\sigma_\cev^2 T}\right)
= O(k^{2\betac + 1}) \qtext{as} k \downarrow 0.
\end{equation}
Finally, the small-strike and small-time asymptotics are 
\begin{equation} \label{eq:cev_asymp}
P_\cev(k) \sim (\text{const.})\, \; k\, T^{1-\frac{1}{2\betac}} \exp\left(-\frac1{2\betacpow{2}\sigma_\cev^2 T} \right) \qtext{as} k \text{ and } T \;\downarrow\; 0.
\end{equation}

\subsection{Observations and insights} \label{ssec:obs} \noindent
Before we proceed to the explicit derivation, it is possible to postulate the form of the equivalent CEV volatility based on insights and observations. The equivalent BS volatility approximations in Section~\ref{sec:model}, after adopting the correction of \citet{obloj2007fine}, can be cast into the following generic form:
\begin{equation} \label{eq:breakdown}
\frac{\sigma_\bs}{\alpha} = \frac{q_\bs}{q}\, H(z)\, \left(\,1 + h_\bs\,T\,\right) \qtext{where} h_\bs = O(\alpha^2) + O(\alpha\vov) + O(\vov^2).
\end{equation}
In Eq.~(\ref{eq:hagan-bs}), for example, the breakdown of $h_\bs$ is obvious. In Eq.~(\ref{eq:paulot-bs}), $A_1$, $A_3$, and $A_2$ are recognized as the terms corresponding to $O(\alpha^2)$, $O(\alpha\vov)$, and $O(\vov^2)$, respectively, based on their values in the limit $k\rightarrow 1$.

To understand the roles of the parts of Eq.~(\ref{eq:breakdown}) better, let us consider the limit $\vov\downarrow 0$. At this limit, the SABR model converges to the CEV model with $\sigma_\cev=\alpha$ and, therefore, Eq.~(\ref{eq:breakdown}) plays the role of converting CEV volatility, $\sigma_\cev=\alpha$, to BS volatility $\sigma_\bs$:
\begin{equation} \label{eq:step1}
\frac{\sigma_\bs}{\sigma_\cev} = \frac{q_\bs}{q} \left(1+ O(\alpha^2)\, T\right).
\end{equation}
Indeed, this form is close to the well-known local volatility conversion of \citet{hagan1999equiv}. However, it is only an approximation; the converted $\sigma_\bs$ does not exactly reproduce the CEV option value for volatility $\sigma_\cev$. Since the next order term is $O(\alpha^4)\, T^2$, approximation quality is expected to be poor for large initial volatility, $\alpha\sqrt{T}>1$.

To preserve the CEV model limit in the volatility approximation, we naturally consider the equivalent CEV volatility. As we recognize that $q_\bs/q$ and $O(\alpha^2)$ are purely involved in the conversion between the CEV and BS models, we do not expect these two terms to appear in the expression of CEV volatility. Dividing Eq.~\eqref{eq:breakdown} by Eq.~\eqref{eq:step1} on each side of the equation, we can factor out these terms and obtain the expected CEV volatility up to $O(T)$ in the form of
\begin{equation} \label{eq:step2}
\frac{\sigma_\cev}{\alpha} = H(z) \big(\,1 + h_\cev\,T\,\big) \qtext{where} h_\cev = O(\alpha\vov) + O(\vov^2).
\end{equation}
This form yields the desired limit, $\sigma_\cev \rightarrow \alpha$ as $\vov\downarrow 0$, owing to the absence of the $O(\alpha^2)$ term and the limit $H\rightarrow 1$ as $z\rightarrow 0$. Since the multiplication of Eqs.~(\ref{eq:step1}) and (\ref{eq:step2}) up to $O(T)$ conversely yields Eq. (\ref{eq:breakdown}), the BS volatility approximation is understood as a two-step conversion: (i) from the SABR model to CEV volatility and (ii) then to BS volatility. Therefore, the CEV volatility approximation in the form above is expected to be more accurate because the second approximation step is omitted. In particular, the advantage of the CEV volatility approach over the BS volatility approach becomes clear for $\alpha\sqrt{T}>1$ (when the quality of the second approximation is poor).

Indeed, we can already appreciate the postulated form for the two special cases of $\beta=0$ and 1. The equivalent normal volatility for the normal SABR model ($\beta=0$), Eq.~(\ref{eq:norm}), follows the form. When $\beta=1$, the equivalent BS volatilities, Eqs.~\eqref{eq:hagan-bs}, \eqref{eq:hklw}, and \eqref{eq:paulot-bs}, all turn into the expected form since $q=q_\bs$ and $\betac=0$. In the next two subsections, we generalize to $0<\beta<1$ by providing explicit derivations.

\subsection{Equivalent CEV volatility based on Hagan's approach} \noindent
We present our first CEV volatility approximation. While we similarly follow the derivation of \citet{hagan2002sabr}, we use the improved normal volatility, Eq.~\eqref{eq:hagan-new}, instead of Eq.~\eqref{eq:hagan-norm}.
\begin{thm} \label{th:cev1}
	The equivalent CEV volatility of the SABR model up to $O(T)$ is given by
\begin{subequations} \label{eq:hagan-cev}
\begin{equation}
\frac{\sigma_\cev}{\alpha} = H(z)\, \left(1 + h_\cev\, T\right),
\end{equation}
where $H(z)$ and $z$ are defined in Eqs.~\eqref{eq:Hxv} and \eqref{eq:zq}, respectively and
\begin{equation} \label{eq:hagan-cev-h}
h_\cev = \frac{\rho}{4} \frac{k^\beta-1}{k-1} \alpha\vov + \frac{2-3\rho^2}{24} \vov^2.
\end{equation}
\end{subequations}
\end{thm}

\begin{pf}
We derive the equivalent CEV volatility in a manner similar to obtaining the equivalent BS volatility in Section~\ref{ssec:hklw}. We equate the normal volatilities for the SABR and CEV models from Eq.~\eqref{eq:hagan-new} as follows:
$$
\sigma_\norm = \alpha\, \frac{q_\norm}{q}\, H(z)\, (1 + h_\norm\, T)
= \sigma_\cev\, \frac{q_\norm}{q} \, \left(1+ \log \left(k^{\beta/2}\frac{q}{q_\norm}\right) \frac{\sigma_\cev^2}{q^2} T\right).
$$
Then, we solve for $\sigma_\cev$ up to $O(T)$ and obtain the result:
\begin{equation}
h_\cev \approx h_\norm - \log \left(k^{\beta/2}\frac{q}{q_\norm}\right) \frac{\sigma_\cev^2}{q^2} \approx \frac{\rho}{4} \frac{k^\beta-1}{k-1} \alpha\vov + \frac{2-3\rho^2}{24} \vov^2,
\end{equation}
where we use $\sigma_\cev \approx \alpha$ from $H=1$ when $k=1$. $\square$
\end{pf}
This approximation indeed meets our expectation, Eq.~\eqref{eq:step2}. Consequently, the SABR model converges to the CEV model; $\sigma_\cev \rightarrow \alpha$ at all $k$ as $\vov \downarrow 0$.

\subsection{Equivalent CEV volatility based on Paulot's approach} \noindent
The equivalent CEV volatility is discussed in \citet[\S 3.6]{paulot2015asym} as an alternative to the equivalent BS volatility. For the CEV model with volatility $\sigma_\cev$, the coefficients of the option value expansion, Eq.~\eqref{eq:optval}, are given by~\citep[p~52]{paulot2015asym}
\begin{equation} \label{eq:cev_ov}
\sigma = \sigma_\cev, \quad
d_\cev = q \;\left(=\frac{z^{\betac} - 1}{\betac}\right), \quad 
e_\cev = \log k^{\beta/2}.
\end{equation}
This is a special case of Eq.~\eqref{eq:sabr_ov} with $\vov\downarrow 0$ and $\alpha=\sigma_\cev$. In turn, Eq.~\eqref{eq:bs_ov} is a special case of Eq.~\eqref{eq:cev_ov} for $\beta\rightarrow 1$ ($\betac\rightarrow 0$). Based on this option value expansion, our second CEV volatility approximation is given below.

\begin{thm} \label{th:cev2}
The equivalent CEV volatility from \citet{paulot2015asym} is simplified to 
\begin{equation} \label{eq:paulot-cev}
\frac{\sigma_\cev}{\alpha} = H(z)\, \left(1 + h_\cev\, T\right) \qtext{where} h_\cev = H(z)^2 \left( A_2 + A_3 \right),
\end{equation}
where $H(z)$ and $z$ are defined in Eqs.~\eqref{eq:Hxv} and \eqref{eq:zq}, respectively and $A_2$ and $A_3$ are given by Eq.~\eqref{eq:paulot-bs-h23}.
\end{thm}

\begin{pf}
Using the expansion in $T$, $\sigma_\cev = \sigma_{\cev,0} (1+ h_\cev T)$, we sequentially obtain the two terms as
$$ \frac{\sigma_{\cev,0}}{\alpha} = \frac{d_\cev}{d_\sabr} = H \\
$$
and
$$ h_\cev = \frac{\alpha^2}{d_\sabr^2}\left(e_\sabr - e_\cev + \log\left(\frac{\alpha}{\sigma_{\cev,0}}\right)\right) = H^2(A_2 + A_3).\quad \square
$$
\end{pf}
The second CEV approximation is compared with the first one in Theorem~\ref{th:cev1} in the same way that the BS volatility of \citet{paulot2015asym} is compared with that of \citet{hagan2002sabr}. The two CEV volatility approximations have the same leading order term, $H$, and their first-order terms, $h_\cev$, also converge to the same value at the money: 
$$ h_\cev \rightarrow \frac{\rho\beta}{4} \alpha\vov + \frac{2-3\rho^2}{24} \vov^2 \qtext{as} k\rightarrow 1.
$$
Similarly, they differ at out-of-the-money strike prices and this difference is expected to be pronounced if $\vov/\alpha$ is large. The second CEV volatility is more accurate and is valid for a wider region of strike prices than the first one because the first-order term, $h_\cev$, is obtained without approximating the dependency on the strike price.

\subsection{Probability mass at zero implied from the CEV volatility approximations} 
\label{ssec:mass} \noindent
We examine the mass at zero and low-strike smile of the two CEV approximations in Theorems~\ref{th:cev1} and \ref{th:cev2}. We show that, unlike the BS-based approximations, the CEV-based approximations can properly imply the probability mass at the origin. We start with the observation that the equivalent CEV volatility has a finite value at $k=0$.
\begin{rem} \label{rem}
	For $0<\beta<1$, the CEV approximations in Theorems~\ref{th:cev1} and \ref{th:cev2} have finite CEV volatilities at a zero strike: 
	$$ \frac{\sigma_\cev^{k=0}}{\alpha} = H\left(-\xi\right)(1+h_\cev^{k=0}\, T) = \frac{\xi\,(1+h_\cev^{k=0}\, T)}{\log \left(\frac{\sqrt{1-2\rho\xi + \xi^2} +\xi -\rho}{1-\rho}\right)} \qtext{for} \xi = \frac{\vov}{\betac\alpha},
	$$
	where the first-order term at $k=0$, $h_\cev^{k=0}$, is finite. If $\rho=0$, the expression is further simplified to
	$$ \frac{\sigma_\cev^{k=0}}{\alpha} = \frac{\xi}{\asinh (\xi)}(1+h_\cev^{k=0}\, T).
	$$
\end{rem}
Here, the CEV volatilities at $k=0$ are evaluated with $z|_{k=0}=-\xi$. In Theorem~\ref{th:cev1}, $h_\cev^{k=0}$ is trivially given by
$$h_\cev^{k=0} = \frac{\rho}{4} \alpha\vov + \frac{2-3\rho^2}{24} \vov^2.
$$
In Theorem~\ref{th:cev2}, a minor complication arises from the evaluation of $G(t_2)$ in Eq.~\eqref{eq:G} at $k=0$ because the argument of log, $\rho + (\eta-\rhoc)t_2 + \sqrt{1-\eta^2}$, approaches zero as $\eta\downarrow 0$. However, this is a removable singularity in the form of $x\log x \rightarrow 0$ as $x\downarrow 0$. Therefore, $G(t_2) = \atan(t_2)$ at $k=0$. 
Now, we use the CEV volatility at $k=0$ to derive the probability mass at zero.
\begin{thm} \label{th:mass}
	For $0<\beta<1$, the mass at zero implied from the CEV approximations (Theorems~\ref{th:cev1} and \ref{th:cev2}) is given by 
	$$
	M_T = \mathbb{P}(f_T = 0) = \bar{\Gamma}\left(\frac{1}{2\betacpow{2}(\sigma_\cev^{k=0})^2\, T};\,\frac{1}{2\betac}\right),
	$$ 
	where $\bar{\Gamma}(x;a)$ is defined in Eq.~\eqref{eq:gammasf}.
\end{thm}
\begin{pf} 
	Given that the equivalent CEV volatility is bounded near $k=0$ and the asymptotic behavior in Eq.~\eqref{eq:cev_2nd}, the small-strike put price from the CEV volatility approximation is similar to that of the CEV model Eq.~\eqref{eq:cev_lowk}:
	\begin{equation} \label{eq:sabr_smallk} 
	P_\sabr(k) \sim k\, \bar{\Gamma}\left(\frac{1}{2\betacpow{2}(\sigma_\cev^{k=0})^2 T};\,\frac{1}{2\betac} \right) \qtext{as} k \downarrow 0.
	\end{equation}
	The mass at zero follows from $M_T = \lim_{k\downarrow 0} P(k)/k$ in Eq.~\eqref{eq:bound}. $\square$
\end{pf}
The numerical experiments in Section~\ref{ssec:num-mass} will show that the above approximation to the mass at zero is accurate. Therefore, Theorem~\ref{th:mass} serves as an alternative to existing methods. Unlike \citet{gulisashvili2018mass}, our method can be used for correlated cases. While \citet{yang2018survival} require a two-dimensional numerical integration for correlated cases, our method is in a closed-form.

Given that $0 < M_T < 1$ under the CEV approximation, we expect that the small-strike smile, when converted to BS volatility, is consistent with Eq.~\eqref{eq:demarco}. Although the result of \citet{demarco2017shapes} is model independent, they use the CEV model as a benchmark for the numerical test because the CEV model is a rare case that renders the exact arbitrage-free option prices and mass at zero. As argued earlier, our CEV approximations are expected to have a lower boundary of arbitrage in the small-strike region.

The mass at zero in Theorem~\ref{th:mass} is also consistent with the findings of \citet{chen2019feeling}. With the absorbing boundary condition at the origin, \citet[Theorem 2.1]{chen2019feeling} proves that a positive constant $T_0$ (depending on the SABR parameters) exists such that
$$ \limsup_{T\downarrow 0}\;T \log M_T \le -T_0,
$$
thereby characterizing the asymptotics of the \textit{not-feeling-boundary} principle as $M_T = O(e^{-T_0/T})$ as $T\downarrow 0$. Theorem~\ref{th:mass} not only is consistent with the asymptotics, but also provides a closed-form expression for $T_0$.
\begin{cor} \label{cor}
In the mass at zero in Theorem~\ref{th:mass}, the time scale of the exponential decay, $T_0$, is given by a closed form:
$$ T_0 = - \lim_{T\downarrow 0}\, T \log M_T = \frac{1}{2\betac^2\alpha^2\xi^2}\log^2 \left(\frac{\sqrt{1-2\rho\xi + \xi^2} +\xi -\rho}{1-\rho}\right) \qtext{for} \xi = \frac{\vov}{\betac\alpha}.$$
\end{cor}
\begin{pf}
From Theorem~\ref{th:mass} and the leading order asymptotics, $\bar{\Gamma}(x;a) \sim x^{a-1} e^{-x} / \Gamma(a)$,
$$ \log M_T = \log \bar{\Gamma}\left( \frac{1}{2\betacpow{2}(\sigma_\cev^{k=0})^2 T};\, \frac{1}{2\betac}\right) \sim - \frac{1}{2\betac ^2 \alpha^2 H(-\xi)^2\, T} \qtext{as} T\downarrow 0.
$$
The expression for $T_0$ naturally follows. Because $T_0$ involves only the leading order CEV volatility at $k=0$, it is independent of the choice of the CEV volatility, presented in either Theorem~\ref{th:cev1} or Theorem~\ref{th:cev2}. $\square$
\end{pf}

It should be noted that the results in Theorem~\ref{th:mass} and Corollary~\ref{cor} are \textit{implied} from our CEV volatility approximations rather than explicitly derived from the SABR dynamics. As such, further mathematical (dis)proof is required to show that the results conform to the true behavior of the SABR model. Below we prove that is the case for $\rho=0$. The proof for the general case, however, is beyond the scope of this paper.
The proof for $\rho=0$ is based on the literature on the exponential functional of BM. 
The exponential functional of BM, defined by
\begin{equation} \label{eq:AT}
A^{[\mu]}_T = \int_0^T \exp(2Z_t + 2\mu t) dt  \qtext{and} A_T = A^{[0]}_T, 
\end{equation}
is a heavily studied topic in stochastic analysis. See \citet{matsuyor2005exp1,matsuyor2005exp2} for an extensive review. Originally, it was inspired by the pricing of continuously monitored Asian options. In the context of the SABR model, however, the integrated variance is related to the functional by
$$ \int_0^T (\alpha\hat{\sigma}_t)^2 dt \;\distequal \left(\frac{\alpha}{\vov}\right)^2 A^{[-1/2]}_{\vov^2 T}.
$$
When $\rho=0$, conditional on the path of $\hat{\sigma}_t$ over $0\le t \le T$, the forward price $f_T$ under the SABR model is distributed according to the CEV model with variance $\sigma_\cev^2 T$ replaced by the integrated variance. Therefore, the option price can be expressed as the expectation of the CEV option price over stochastic variance $(\alpha/\vov)^2 A^{[-1/2]}_{\vov^2 T}$~\citep{islah2009sabr-lmm}. The mass at zero for $\rho=0$ is similarly expressed as~\citep{gulisashvili2018mass}
\begin{equation} \label{eq:MTMC}
M_T = E\left[\bar{\Gamma}\left(\frac{\xi^2}{2 A^{[-1/2]}_{\vov^2 T}};\,\frac{1}{2\betac} \right)\right] \qtext{for} \xi = \frac{\vov}{\betac\alpha}.
\end{equation}

\begin{thm} \label{th:T0}
	When $\rho=0$ and $0<\beta<1$, the decay time scale $T_0$ from Corollary~\ref{cor}, simplified to
	\begin{equation*} 
	T_0 = -\lim_{T\downarrow 0} T \log M_T = \frac{1}{2\betac^2\alpha^2} \frac{\asinh(\xi)^2}{\xi^2} \qtext{for} \xi = \frac{\vov}{\betac\alpha},
	\end{equation*}
	is consistent with the true SABR dynamics.
\end{thm}
\begin{pf}
We begin the proof with the Laplace transform of $1/A_t$ from \citet[Theorem~5.6]{matsuyor2005exp1}:
\begin{equation} \label{eq:laplace}
	E\left[ \exp\left(-\frac{\xi^2}{2 A_t}\right) \right] = \frac{1}{\sqrt{2\pi}}\int_{-\infty}^{\infty} \!dx\exp\left(-\frac{\acosh((\xi^2/2) e^{-x} +\cosh(x))^2}{2t}\right).
\end{equation}
By applying Laplace's method, we show that
$$
- \lim_{t\downarrow 0} t\, \log E\left[\exp\left(-\frac{\xi^2}{2 A_t}\right) \right] = - \lim_{t\downarrow 0} t\, \log\left[(\text{const.})\sqrt{t}\, \exp\left( -\frac{\acosh(\sqrt{1+\xi^2}\,)^2}{2t}\right)\right] = \frac{\asinh(\xi)^2}{2},
$$
because the minimum of $(\xi^2/2) e^{-x} +\cosh(x)$ occurs when $e^x = \sqrt{1+\xi^2}$. It can be also derived from an alternative expression of the Laplace transform~\citep[Eq.~(3.108)]{antonov2019modern}.
With Girsanov's theorem and the asymptotic expansion of $\bar{\Gamma}(x;a)$ in Eq.~\eqref{eq:gammasf_asymp}, $M_T$ can be expressed as
\begin{align*}
M_T &= E\left[ \bar{\Gamma}\left(\frac{\xi^2}{2 A^{[\mu]}_t};\,a \right) \right]
= E\left[ \bar{\Gamma}\left(\frac{\xi^2}{2 A_t};\,a \right) \exp\left(\mu Z_t - \frac{\mu^2 t}{2}\right) \right]\\
&= \frac{1}{\Gamma(a)}E\left[ \exp\left(-\frac{\xi^2}{2 A_t} + (a-1)\log(A_t) + O(A_t) + \mu Z_t - \frac{\mu^2 t}{2}\right) \right],
\end{align*}
where $\mu=-1/2$, $a = 1/(2\betac)$, and $t=\vov^2 T$. 
Since $A_t = O(t)$ as $t\downarrow 0$, the first term, $-\xi^2/(2 A_t)$, of the exponent asymptotically dominates the rest as $t\downarrow 0$. 
The small-time asymptotics of $M_T$ should have the same leading order term of Eq.~\eqref{eq:laplace}. Therefore, the result follows as 
$$ T_0 = - \lim_{T\downarrow 0}  T \log M_T
= - \lim_{T\downarrow 0} T\, \log E\left[\exp\left(-\frac{\xi^2}{2 A_{\vov^2 T}}\right) \right] = \frac{1}{2\betac^2\alpha^2}\frac{\asinh(\xi)^2}{\xi^2}. \quad\square
$$
\end{pf}

Note that two special cases of the expectation in similar forms are exactly known for $t>0$~\citep[Corollary 4.6]{matsuyor2005exp1}:
\begin{gather*}
E\left[ \frac{1}{\sqrt{A_t}} \exp\left(-\frac{\xi^2}{2 A_t}\right) \right] =
\frac1{\sqrt{t\,(1+\xi^2)}} \exp\left(-\frac{\asinh(\xi)^2}{2 t}\right), \\
E\left[ \frac{1}{\sqrt{A^{[1]}_t}} \exp\left(-\frac{\xi^2}{2 A^{[1]}_t}\right) \right] =
\frac{e^{-t/2}}{\sqrt{t}} \exp\left(-\frac{\asinh(\xi)^2}{2 t}\right).
\end{gather*}
and both satisfy the small-time limit, $-\lim_{t\downarrow 0} t \log E[\;\cdot\;] = \asinh(\xi)^2 / 2$.

We comment on (in)consistencies between the approximation methods for the mass at zero. In the $\vov\downarrow 0$ limit, the mass at zero from the three methods, \citet{yang2018survival}, \citet{gulisashvili2018mass}, and Theorem~\ref{th:mass}, all converges to that of the CEV model with $\sigma_\cev=\alpha$, which is consisent with the intution. In Theorem~\ref{th:mass}, this is the case because $\sigma_\cev^{k=0}\rightarrow\alpha$ as $\vov\downarrow 0$. We note, however, the difference in decay time scales (i.e., the $T\downarrow0$ limit) between Theorem~\ref{th:T0} and \citet{yang2018survival}. When $\rho=0$, the mass at zero from \citet[Eqs.~(8) and (9)]{yang2018survival} reduces to that of the CEV model even though $\vov$ is not small. That implies $-T\log M_T \rightarrow 1/(2\betac^2\alpha^2)$ as $T\downarrow 0$. 
However, this is significantly different from $T_0$ in Theorem~\ref{th:T0} if $\xi = \vov/(\betac\alpha)$ is in the order of 1. The numerical experiment in Section~\ref{ssec:num-mass} (Figure~\ref{fig:set1_m0}) is in favor of our asymptotic time scale.

The success of our CEV approximations in estimating the mass at zero is surprising considering that the boundary condition has not been explicitly considered in the derivation of the CEV approximations. Below are our insights on how this happens. We argue that the CEV-based approach is effective because it works as a control variate method. Recall that the CEV volatility based on \citet{paulot2015asym} in Theorem~\ref{th:cev2} is derived by matching the small-time option values between the CEV and SABR models: 
\begin{equation} \label{eq:ov_equate}
\frac{\sigma_\cev^3 T^{3/2}\,k^{\beta/2}}{q^2} \exp\left(-\frac{q^2}{2\sigma_\cev^2 T} + \cdots\right)
= \frac{\alpha^3 T^{3/2}\,k^{\beta/2}}{(q/H)^2}\; \exp\left(-\frac{(q/H)^2}{2\alpha^2 T} + \cdots\right),
\end{equation}
where the left-hand side from Eqs.~\eqref{eq:optval} and \eqref{eq:cev_ov} is for the CEV model and the right-hand side from Eqs.~\eqref{eq:optval} and \eqref{eq:sabr_ov} is for the SABR model. 
The asymptotics on each side are originally intended to hold in the small-time limit near $k=1$ in general. As such, they are not accurate near $k=0$. The correct CEV price asymptotics in Eq.~\eqref{eq:cev_asymp} are different from the CEV value on the left-hand side in terms of the powers of $k$ and $T$ in the prefactor. However, the incorrect term $T^{3/2}k^{\beta/2}$ also arises on the right-hand side for the SABR model and they just cancel out. The choice of the CEV model also makes the exponents of the two sides closer (e.g., $q^2$ versus $(q/H)^2$). As a result, the CEV approach prevents the equivalent volatility from diverging to infinity as $k\downarrow 0$. Finally, the SABR price asymptotics, Eq.~\eqref{eq:sabr_smallk}, obtained via the equivalent CEV volatility are similar to those of the CEV model, Eq.~\eqref{eq:cev_lowk}, which correctly carries the absorbing boundary condition. Notably, the output, Eq.~\eqref{eq:sabr_smallk}, is correct even though the input, Eq.~\eqref{eq:ov_equate}, is incorrect because the CEV approach works as a control variate method. 

\section{Numerical results} \label{sec:num} \noindent
In this section, we numerically test the two CEV-based approximations in comparison to existing BS-based approximations.\footnote{The Python code used in this study can be found at \url{https://github.com/PyFE/PyfengForPapers}.} For easier reference, the methods are labeled as follows:
\begin{itemize}
	\item \textbf{BS-A}: Eq.~\eqref{eq:hklw}, the HKLW formula of \citet{hagan2002sabr}.
	\item \textbf{BS-B}: Eq.~\eqref{eq:paulot-bs}, \citet{paulot2015asym}'s equivalent BS volatility of order $O(T)$.
	\item \textbf{BS-C}: \citet[\S 5.4]{lorig2017lsv}'s equivalent BS volatility up to $O(T^3)$
	\item \textbf{DMHJ}: Eq.~\eqref{eq:demarco}, the small-strike BS volatility smile determined by the mass at zero~\citep{demarco2017shapes}
	\item \textbf{CEV-A}: Eq.~\eqref{eq:hagan-cev} in Theorem~\ref{th:cev1}, the equivalent CEV volatility based on \citet{hagan2002sabr}.
	\item \textbf{CEV-B}: Eq.~\eqref{eq:paulot-cev} in Theorem~\ref{th:cev2}, the equivalent CEV volatility based on \citet{paulot2015asym}.
\end{itemize}
The three BS-based approximations are good representatives of existing methods. Let us make several comments on the selection of the methods. In BS-A, we do not adopt \citet{obloj2007fine}'s leading order correction since the HKLW formula is widely used already. In BS-B, we do not use the $O(T^2)$ term because it is not in a closed form. The naive CEV approximation of \citet{yang2017cev-sabr} is not included because it has no dependency on $\vov$ and $\rho$. Similarly, the price approximation of \citet{jordan2011vol} is not included as it works only for $\rho=0$. Although it is not reviewed in Section \ref{sec:model}, we include the $O(T^3)$ approximation of \citet{lorig2017lsv}, which is labeled BS-C. In \citet{lorig2017lsv}, the higher-order result is achieved at the expense of the lower accuracy for off-the-money strike; the approximation is valid only for near-the-money strike prices because the dependency on $k$ is expressed in terms of the expansions of $\log k$ in all orders of time. This is in contrast to BS-A and BS-B, in which the accuracy over all $k$ is maintained to a certain degree through $H(z)$ in the leading order term\footnote{The leading and first-order terms of BS-C coincide with those of BS-A and BS-B at $k=1$.}. For this reason, the approximation quality of BS-C may deteriorate faster than that of BS-A or BS-B as $k$ moves away from one. Therefore, it is of additional interest to numerically compare the three BS-based methods. The DMHJ asymptotics is not self-contained, as it needs the externally estimated mass at zero and are only valid for $k<1$. However, they provide a useful reference as a model-free volatility smile at small strikes. We thus evaluate DMHJ with the two values of the mass at zero; one estimated from CEV-A or CEV-B and the other the true value.

We also compare the following methods to estimate the mass at zero:  
\begin{itemize}
	\item \textbf{GHJ-LN}: The moment-matched lognormal approximation of \citet{gulisashvili2018mass} for $\rho=0$ in small-time limit.
	\item \textbf{YW}: \citet[Eqs.~(8) and (9)]{yang2018survival}.
	\item \textbf{MC}: Eq.~\eqref{eq:MTMC} with Monte-Carlo simulated $A^{[-1/2]}_{\vov^2 T}$ for $\rho=0$.
	\item \textbf{CEV-A}: Theorem~\ref{th:mass} with the zero strike CEV volatility from Theorem~\ref{th:cev1}.
	\item \textbf{CEV-B}: Theorem~\ref{th:mass} with the zero strike CEV volatility from Theorem~\ref{th:cev2}.
\end{itemize}
In GHJ-LN, the integrated variance is approximated by a lognormal distribution by matching to the first and second moments. Thus, the expectation in Eq.~\eqref{eq:MTMC} is computed as a one-dimensional integration. While the reference does not specify the numerical method, we use the Gauss--Hermite quadrature for efficient integration with fast convergence~\citep{choi2021note}.
In YW, the mass at zero is expressed as an expansion consist of the $O(1)$ and $O(\vov\sqrt{T})$ terms. As mentioned earlier, when $\rho=0$, the first order term vanishes and the leading order term is reduced to that of the CEV model. For accuracy benchmarking, we also implement the Monte-Carlo (MC) method. We draw the random values of $A^{[-1/2]}_{\vov^2 T}$ by simulating $\hat{\sigma}_t$ on a discretized time grid and integrating $\hat{\sigma}_t^2$ by Simpson's rule for quadrature. We generate 40,000 paths with 20 time steps between $t=0$ and $T$.

Table~\ref{tab:param} shows the parameter sets used for numerical test. Sets~1--3 are for the volatility approximation (Section~\ref{ssec:num-vol}) and the arbitrage boundary (Section~\ref{ssec:num-smallk}), while Sets~1, 4, and 5 are for the mass at zero (Section~\ref{ssec:num-mass}).

\subsection{Volatility approximation error} \noindent
\label{ssec:num-vol}
First, we test the accuracy of the analytic approximations using Sets 1--3. The three parameter sets are carefully chosen for comparing the relative strength of various methods. Those sets have been used in past studies~\citep{vonsydow2019benchop,cai2017sabr,antonov2019modern} and the exact option prices are available through numerical means (e.g., the finite difference method and the Monte-Carlo simulation). These exact values are additionally verified within reasonable accuracy using the continuous-time Markov chain codes\footnote{\url{https://github.com/jkirkby3/PROJ_Option_Pricing_Matlab}} adopted by \citet{cui2018ctmc-sabr,cui2019fullfledged}\footnote{The result for Set~2 is also reported in \citet[Table~2]{cui2018ctmc-sabr} and \citet[Table~1]{cui2019fullfledged}.}. The table also displays $\alpha\sqrt{T}$ and $\vov\sqrt{T}$ as references to the performance of each method. The accuracy of the approximations tends to deteriorate as the two variables become larger. The ratio $\vov/\alpha$ also indicates how much Paulot's refinement is noticeable.

Tables~\ref{tab:set1}--\ref{tab:set3} respectively compare the approximation accuracy for the three parameter sets. Each table shows the standardized BS volatility error, $(\sigma_\bs - \sigma_\bs^\text{exact})/\alpha$, where $\sigma_\bs$ is the implied BS volatility from the analytic approximation methods\footnote{In the BS-based methods, $\sigma_\bs$ is directly obtained from the volatility formulas, whereas in the CEV-based methods, $\sigma_\bs$ is converted from the CEV option prices in Eqs.~\eqref{eq:cev_call}--\eqref{eq:cev_put}.} and $\sigma_\bs^\text{exact}$ is the exact BS volatility reported in the literature. The $\sigma_\bs^\text{exact}$ value and corresponding call option price (in the original scale) are also provided in the tables for reference.

Table~\ref{tab:set1} shows that all the methods are accurate for Set~1 with an error below 0.02. This is expected from the values of $\vov\sqrt{T}$ and $\alpha\sqrt{T}$ not exceeding one. Among the methods, BS-C is the most accurate because it contains higher-order terms and the tested strikes are near the money. The two CEV-based methods are more accurate than the first two BS-based methods by a small margin.

Set~2 is often used in the literature~\citep{cai2017sabr,cui2018ctmc-sabr} to reveal the failure of the HKLW formula. Table~\ref{tab:set2} shows that the CEV-based approximations are superior to their BS-based counterparts because of the large value of $\alpha\sqrt{T}=3.257$. Owing to the absence of the $O(\alpha^2)\,T$ term, the CEV-based approximations have a much smaller error when $\alpha\sqrt{T}>1$, whereas both BS-A and BS-B largely over-predict volatility. The difference between BS-A and CEV-A (and between BS-B and CEV-B) is believed to be as the error generated from converting CEV volatility to BS volatility (see Section~\ref{ssec:obs}), and such error can be avoided by adopting CEV volatility instead. In addition, the numerical results are close between BS-A and BS-B and between CEV-A and CEV-B because the small $\vov/\alpha$ ratio makes $z$ small for fixed $k$ and, therefore, makes the type--B methods equipped with \citet{paulot2015asym}'s refinement indistinguishable from the type--A methods. BS-C shows an interesting result. While it is the most accurate at the money ($k=1$) as expected, the error rises in the out-of-the-money region. Notably, the volatility skew (i.e., the slope of $\sigma_\bs$) by BS-C is in the opposite direction to those from the other methods as well as the exact skew.

Set~3 is perhaps the most challenging case because of the large values of $\vov\sqrt{T}$ and $\alpha\sqrt{T}$, which are both higher than one. Table~\ref{tab:set3} shows the results. Although the errors are large overall as expected, CEV-B shows higher accuracy over the other methods except BS-C. In general, the B-type methods perform better than the A-type methods. This grouping of the results is different from the two previous test sets because $z$ is amplified by the large $\vov/\alpha$ ratio and \citet{paulot2015asym}'s refinement dominates the CEV volatility effect. BS-C is the most accurate among the tested methods, although not by a significant amount.

\begin{table} 
	\caption{Parameter sets tested. The parameters super-scripted by * are to be varied from the base values in numerical tests.} \label{tab:param}
	\begin{center}
	\begin{tabular}{|c||c|c|c|c|c|c||c|c|c|} \hline 
Set & $\beta$ & $\sigma_0$ & $\vov$ & $\rho$ & $T$ & $F_0$ & $\alpha\sqrt{T}$ & $\vov\sqrt{T}$ & Results\\ \hline\hline
1 & 0.5 & 0.5 & 0.4 & 0 & $2^*$ & 0.5 & 1.000 & 0.566 & Table~\ref{tab:set1}, Figs.~\ref{fig:set1_m0}, \ref{fig:demarco}(a)\\
2 & 0.3 & 0.4 & 0.6 & 0 & 1 & 0.05 & 3.257 & 0.600 & Table~\ref{tab:set2}, Fig.~\ref{fig:demarco}(b)\\
3 & $0.6^*$ & $0.25^*$ & $0.3^*$ & $-0.2$ & 20 & 1 & 1.118 & 1.342 & Table~\ref{tab:set3}, Figs.~\ref{fig:demarco}(c), \ref{fig:pdf}, \ref{fig:arb}\\
4 & $0.1^*$	& $0.1^*$ & $0.1^*$	& $-0.5$ & 0.1 -- 1.0 & $0.1^*$ & 0.50 -- 1.59 & 0.03 -- 0.10 & Fig.~\ref{fig:set4_m0}, Fig.~\ref{fig:demarco}(d)\\
5 & $0.1^*$	& 0.1 & 0.1	& $0^*$	& 0.5 & 0.1 & 0.562 & 0.071 & Table~\ref{tab:mass} \\ \hline
	\end{tabular}
	\end{center}
\end{table}

\begin{table} 
	\caption{Accuracy of the various approximation methods for \hyperref[tab:param]{Set~1}. The standardized volatility error, $(\sigma_\bs - \sigma_\bs^\text{exact})/\alpha$, is displayed, where $\sigma_\bs$ is the equivalent BS volatility from the analytic approximations and $\sigma_\bs^\text{exact}$ is the exact BS volatility converted from the price reported in \citet[Table~A1]{vonsydow2019benchop}. For the CEV-based methods, $\sigma_\bs$ is converted from $\sigma_\cev$ using the CEV option price, Eq.~(\ref{eq:cev_call}).} \label{tab:set1}
	\begin{center}
	\begin{tabular}{|c|c||c|c|c|c|c||c|c|} \hline
\multicolumn{2}{|c||}{Strike} & \multicolumn{5}{c||}{$(\sigma_\bs - \sigma_\bs^\text{exact})/\alpha$} & $\sigma_\bs^\text{exact}$(\%) & Price \\ \hline
$k$ & $z$ & BS-A & BS-B & BS-C & CEV-A & CEV-B & \multicolumn{2}{c|}{Exact} \\ \hline
0.868 & -0.077 & 0.027 & 0.027 & 0.004 & 0.024 & 0.024 & 74.19 & 0.221383 \\
1 & 0 & 0.024 & 0.024 & 0.003 & 0.022 & 0.022 & 71.67 & 0.193837 \\
1.152 & 0.083 & 0.021 & 0.021 & 0.002 & 0.019 & 0.019 & 69.33 & 0.166241 \\
\hline
	\end{tabular}
	\end{center}
\end{table}

\begin{table} 
	\caption{Accuracy of the various approximation methods for \hyperref[tab:param]{Set~2}. The standardized volatility error, $(\sigma_\bs - \sigma_\bs^\text{exact})/\alpha$, is displayed, where $\sigma_\bs$ is the equivalent BS volatility from the analytic approximations and $\sigma_\bs^\text{exact}$ is the exact BS volatility converted from the price reported in \citet[Table~7]{cai2017sabr}. For the CEV-based methods, $\sigma_\bs$ is converted from $\sigma_\cev$ using the CEV option price, Eq.~(\ref{eq:cev_call}).} \label{tab:set2}
	\begin{center}
		\begin{tabular}{|c|c||c|c|c|c|c||c|c|} \hline
\multicolumn{2}{|c||}{Strike} & \multicolumn{5}{c||}{$(\sigma_\bs - \sigma_\bs^\text{exact})/\alpha$} & $\sigma_\bs^\text{exact}$(\%) & Price \\ \hline
$k$ & $z$ & BS-A & BS-B & BS-C & CEV-A & CEV-B & \multicolumn{2}{c|}{Exact} \\ \hline
0.4 & -0.125 & 1.059 & 1.041 & -0.636 & 0.051 & 0.051 & 292.47 & 0.0456 \\
0.8 & -0.038 & 0.586 & 0.586 & -0.167 & 0.051 & 0.051 & 260.51 & 0.0414 \\
1 & 0 & 0.480 & 0.480 & -0.006 & 0.050 & 0.050 & 249.62 & 0.0394 \\
1.2 & 0.036 & 0.405 & 0.405 & 0.129 & 0.049 & 0.049 & 240.79 & 0.0375 \\
1.6 & 0.103 & 0.308 & 0.307 & 0.349 & 0.047 & 0.047 & 226.38 & 0.0339 \\
2.0 & 0.164 & 0.247 & 0.246 & 0.525 & 0.045 & 0.045 & 215.05 & 0.0306 \\
\hline
		\end{tabular}
	\end{center}
\end{table}

\begin{table} 
\caption{Accuracy of the various approximation methods for \hyperref[tab:param]{Set~3}. The standardized volatility error, $(\sigma_\bs - \sigma_\bs^\text{exact})/\alpha$, is displayed, where $\sigma_\bs$ is the equivalent BS volatility from the analytic approximations and $\sigma_\bs^\text{exact}$ is the exact BS volatility reported in \citet[Table~4.2]{antonov2019modern}. For the CEV-based methods, $\sigma_\bs$ is converted from $\sigma_\cev$ using the CEV option price, Eq.~(\ref{eq:cev_call}).} \label{tab:set3}
\begin{center}
	\begin{tabular}{|c|c||c|c|c|c|c||c|c|} \hline
		\multicolumn{2}{|c||}{Strike} & \multicolumn{5}{c||}{$(\sigma_\bs - \sigma_\bs^\text{exact})/\alpha$} & $\sigma_\bs^\text{exact}$(\%) & Price \\ \hline
		$k$ & $z$ & BS-A & BS-B & BS-C & CEV-A & CEV-B & \multicolumn{2}{c|}{Exact} \\ \hline
		0.1 & -1.806 & 0.710 & 0.592 & -0.463 & 0.597 & 0.435 & 41.22 & 0.9222 \\
		0.4 & -0.921 & 0.365 & 0.298 & -0.046 & 0.353 & 0.278 & 29.73 & 0.7082 \\
		0.8 & -0.256 & 0.226 & 0.213 & -0.117 & 0.224 & 0.209 & 24.18 & 0.4772 \\
		1.0 & 0.000 & 0.195 & 0.195 & -0.138 & 0.194 & 0.194 & 22.73 & 0.3887 \\
		1.2 & 0.227 & 0.178 & 0.184 & -0.151 & 0.177 & 0.183 & 21.81 & 0.3182 \\
		1.6 & 0.621 & 0.169 & 0.169 & -0.159 & 0.168 & 0.169 & 20.97 & 0.2215 \\
		2.0 & 0.959 & 0.174 & 0.162 & -0.145 & 0.175 & 0.161 & 20.81 & 0.1637 \\
		\hline
	\end{tabular}
\end{center}
\end{table}

\subsection{Mass at zero and its small-time asymptotics} \label{ssec:num-mass}

\begin{figure}[ht!]
	\caption{\label{fig:set4_m0} The mass at zero, $M_T$, from Theorem~\ref{th:mass} for \hyperref[tab:param]{Set~4}. The decay ratio, $-T \log M_T$, scaled by its limit $T_0$ (displayed in each subplot title) from Corollary~\ref{cor} is plotted as a function of $T$. The zero-strike CEV volatilities obtained from the CEV-A and CEV-B methods are used to compute $M_T$. The exact values computed with the finite difference method are from \citet[Figure~1]{chen2019feeling}. Subplot (a) uses the base parameter values and the rest use modified parameters (displayed in each subplot title). The axis titles are shown only in (a) for brevity.
	} \vspace{2ex}
	\centering
	\includegraphics[width=0.52\textwidth]{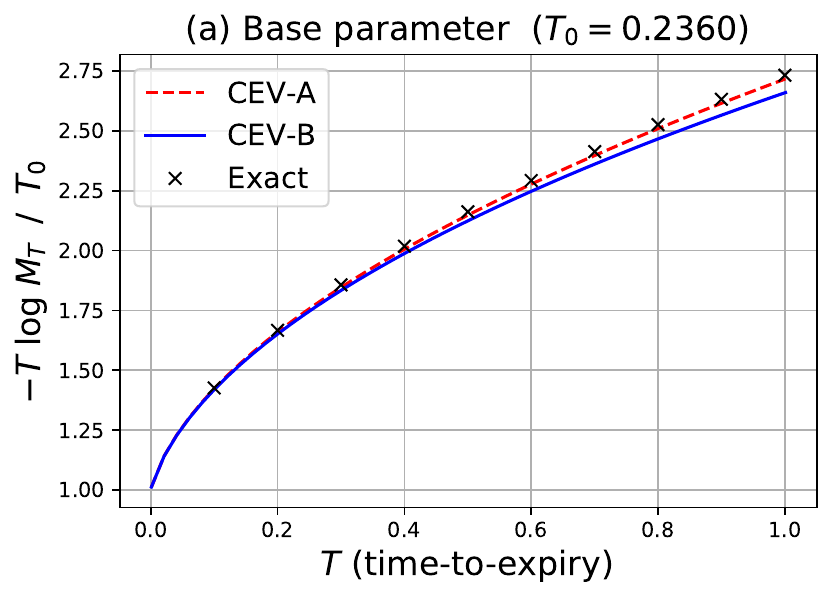}\\ \vspace{1ex}
	\includegraphics[width=0.47\textwidth]{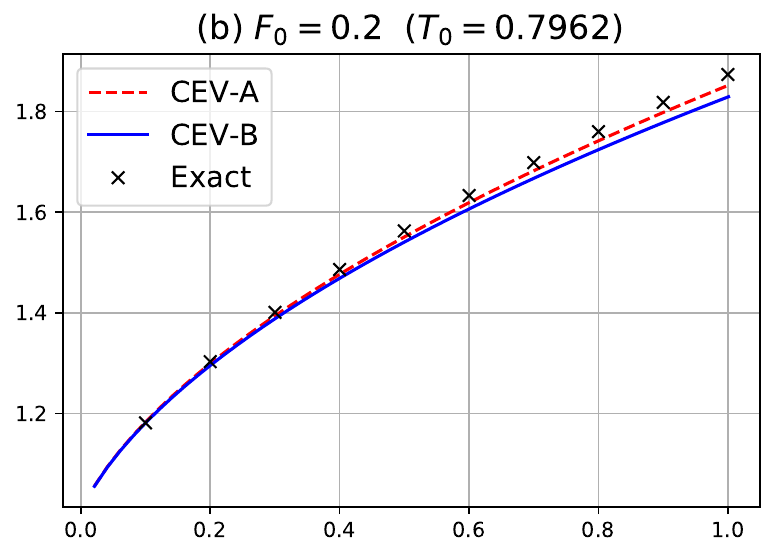}
	\includegraphics[width=0.47\textwidth]{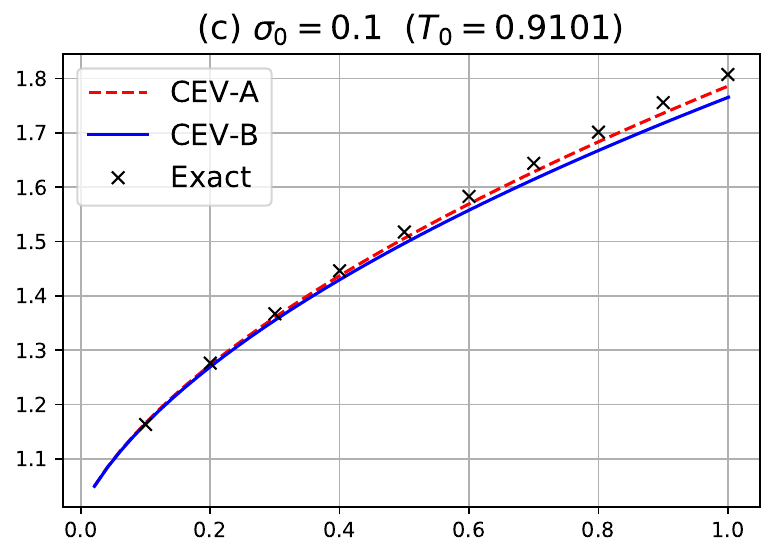}\\ \vspace{1ex}
	\includegraphics[width=0.47\textwidth]{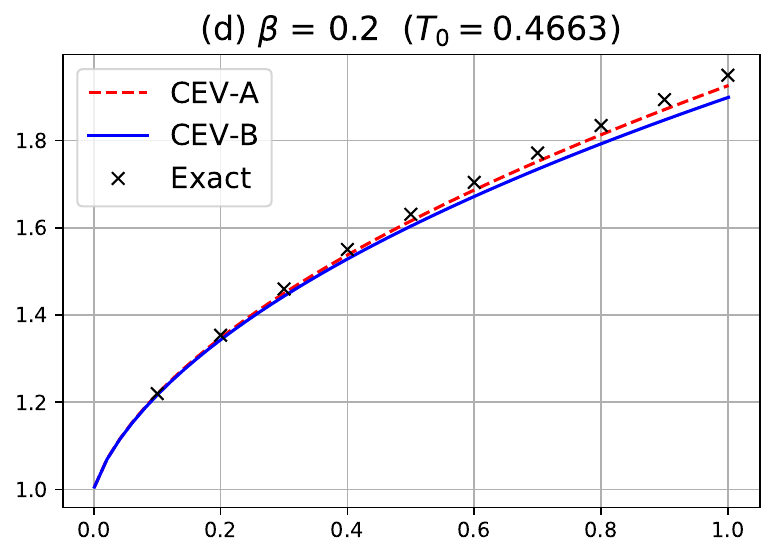}
	\includegraphics[width=0.47\textwidth]{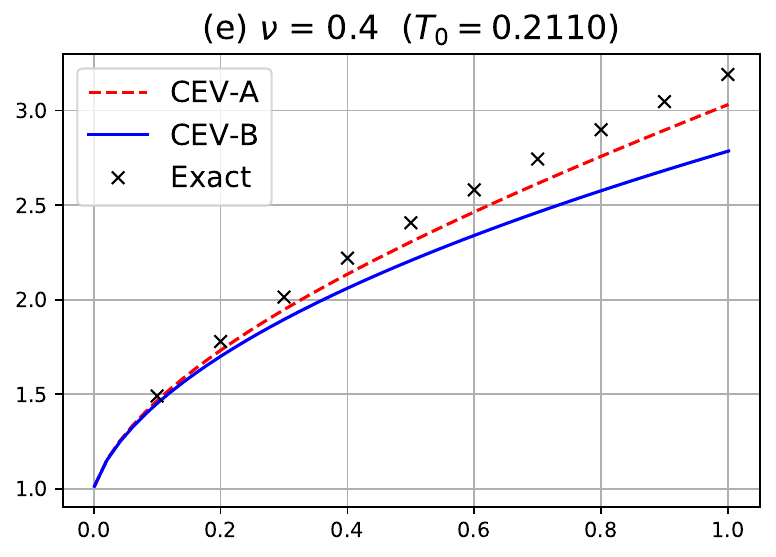}\\ \vspace{2cm}
\end{figure}

\noindent To demonstrate the accuracy of the mass-at-zero approximation in Theorem~\ref{th:mass}, we test with Set~4. \citet[Figure~1]{chen2019feeling} report the mass-at-zero values computed with the finite difference method for Set~4 and several modifications from the base values. Figure~\ref{fig:set4_m0} shows the comparison. To efficiently compare the small-time asymptotics of $M_T$, we plot the ratio of decay time scale, $-T \log M_T / T_0$, as a function of $T$, where $T_0$ is from Corollary~\ref{cor}. If $M_T$ follows Corollary~\ref{cor}, $-T \log M_T / T_0 \rightarrow 1$ as $T\downarrow 0$. Therefore, Corollary~\ref{cor} is verified from the plot.
CEV-A and CEV-B from Theorem~\ref{th:mass} show good agreement with the exact values overall even though it is a correlated case ($\rho=-0.5$). In particular, they are accurate for small $T$, supporting Corollary~\ref{cor}. Among the parameter variations, the increased vol-of-vol in subplot (e) shows the largest error for the same level of $T$. This is consistent with fact that the analytic approximation deteriorates as $\vov\sqrt{T}$ rises. Unexpectedly, CEV-A shows slightly higher accuracy than CEV-B, which may just be a coincidence for this parameter set.

\begin{figure}[ht!]
	\caption{\label{fig:set1_m0} The mass at zero, $M_T$, measured with  various methods for \hyperref[tab:param]{Set~1}. The same data is displayed with two different $y$-axis scales: decay ratio, $-T \log M_T / T_0$ (left) and absolute value, $M_T$ (right). We use $T_0=2.9481$ from Theorem~\ref{th:T0}.} \vspace{2ex}
	\centering
	\includegraphics[width=0.47\textwidth]{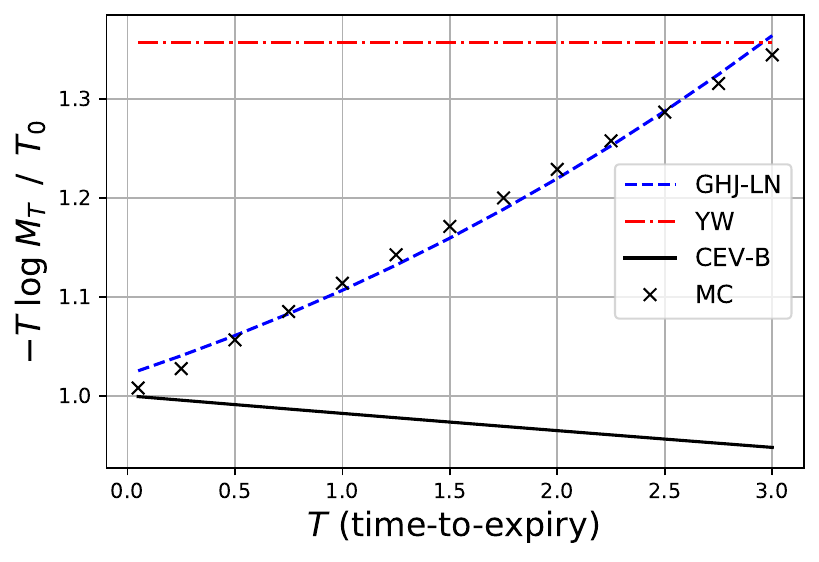}
	\includegraphics[width=0.47\textwidth]{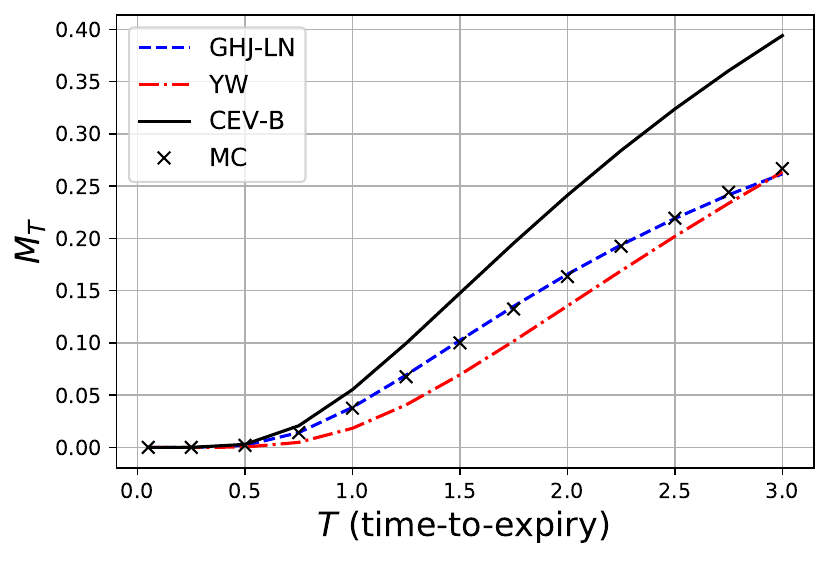}
\end{figure}

With Set~1, we further investigate the exponential decay rate and the validity of Theorem~\ref{th:T0} when $\rho=0$. Since $\rho=0$, the comparison among various methods, including GHJ-LN and MC, is possible. Since $\beta=\betac=0.5$, the mass at zero under YW simplifies to $M_t = e^{-2/(\alpha^2T)}$ and, therefore, the decay time scale ratio is expected to be a constant, $-T\log M_T /T_0 = \xi^2 / \asinh(\xi)^2 = 1.3568$, regardless of $T$. On the contrary, the ratio converges to 1 under our CEV approximations (Theorem~\ref{th:T0}). Figure~\ref{fig:set1_m0} (left) indeed shows that CEV-B is in favor; both MC and GHJ-LN converge to 1 as $T\downarrow 0$. Among the three approximations, GHJ-LN shows the best accuracy over a wide range of $T$ beyond `small' time.

Table~\ref{tab:mass} shows the result for Set~5 and the variations with respect to $\beta$ and $\rho$. The parameter has been tested in \citet[Table~A.5]{yang2018survival} and the values for finite difference (all $\rho$) and YW ($\rho\neq 0$) are from the reference. The finite difference for $\rho\neq 0$ and MC for $\rho=0$ are considered the most accurate method. Although GHJ-LN is only available for $\rho=0$, it again shows the best agreement with MC. Overall, both CEV-B\footnote{The CEV-A results are very close to those of CEV-B.} and YW agree with the benchmark values for all cases.

\begin{table}[ht!]
	\caption{The mass at zero, $M_T$, computed with various methods for \hyperref[tab:param]{Set~5} and the variations on $\beta$ and $\rho$. The finite difference values for all $\rho$ and the YW values for $\rho\neq 0$ are from the survival probability, $1-M_T$, reported in \citet[Table~A.5]{yang2018survival}.}\label{tab:mass}
	\begin{center}
		\begin{tabular}{|c||ccc|cccc|} \hline
			& \multicolumn{3}{c|}{$\beta=0.1$} & $\beta=0.1$ & 0.2 & 0.3 & 0.4 \\ \cline{2-8}
			Method & $\rho=-0.3$ & $-0.2$ & $-0.1$ & \multicolumn{4}{|c|}{$\rho=0$} \\ \hline
			GHJ-LN & & & & 5.66E-2 & 7.97E-3 & 1.50E-4 & 3.20E-8 \\
			MC & & & & 5.66E-2 & 7.97E-3 & 1.50E-4 & 3.25E-8 \\
			CEV-B & 6.22E-2 & 6.05E-2 & 5.88E-2 & 5.70E-2 & 8.09E-3 & 1.55E-4 & 3.48E-8 \\
			YW & 0.0595 & 0.0583 & 0.0572 & 5.61E-2 & 7.64E-3 & 1.24E-4 & 1.33E-8 \\
			Finite Difference & 0.0606 & 0.0593 & 0.0580 & 0.0567 & 0.0081 & 0.0002 & 0.0000 \\
			\hline
		\end{tabular}
	\end{center}
\end{table}

\subsection{Small-strike smile and arbitrage boundary} \noindent
\label{ssec:num-smallk}

\begin{figure}[ht!]
\caption{\label{fig:demarco} The comparison of the low-strike BS volatility smiles for \hyperref[tab:param]{Sets~1--4}. DMHJ uses the mass at zero ($M_T$) implied from CEV-B via Theorem~\ref{th:mass} (the first value in each subplot title), whereas DMHJ-2 uses the true value of $M_T$ (the second value). Exact volatilities for (a)--(c) are from Tables~\ref{tab:set1}--\ref{tab:set3} respectively. For (d), $\vov=0.4$ and $T=1$ are used with the base parameters.} \vspace{2ex}
	\centering
	\includegraphics[width=0.49\textwidth]{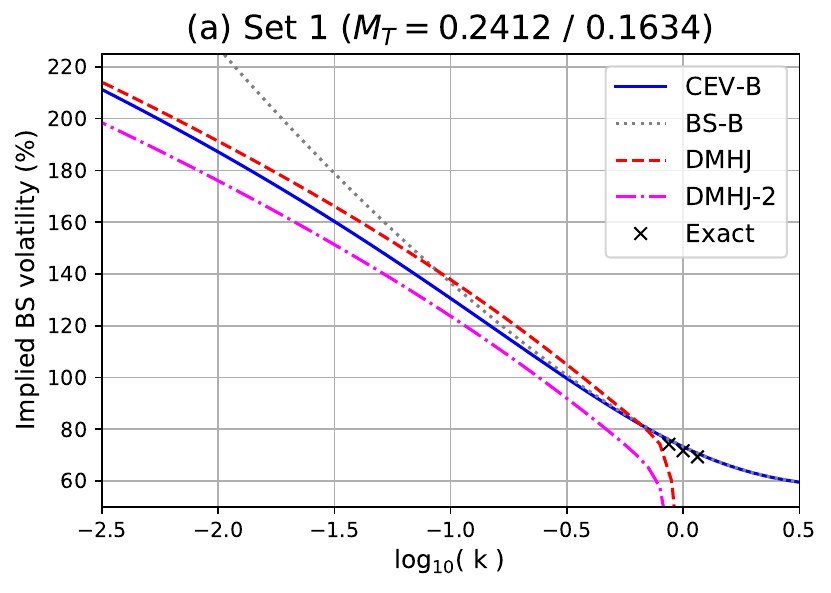}
	\includegraphics[width=0.49\textwidth]{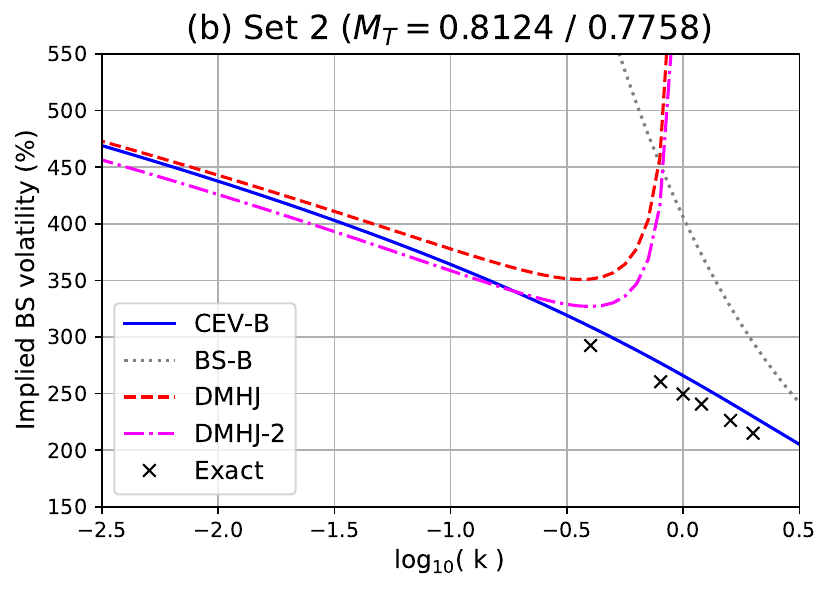}\\ \vspace{2ex}
	\includegraphics[width=0.49\textwidth]{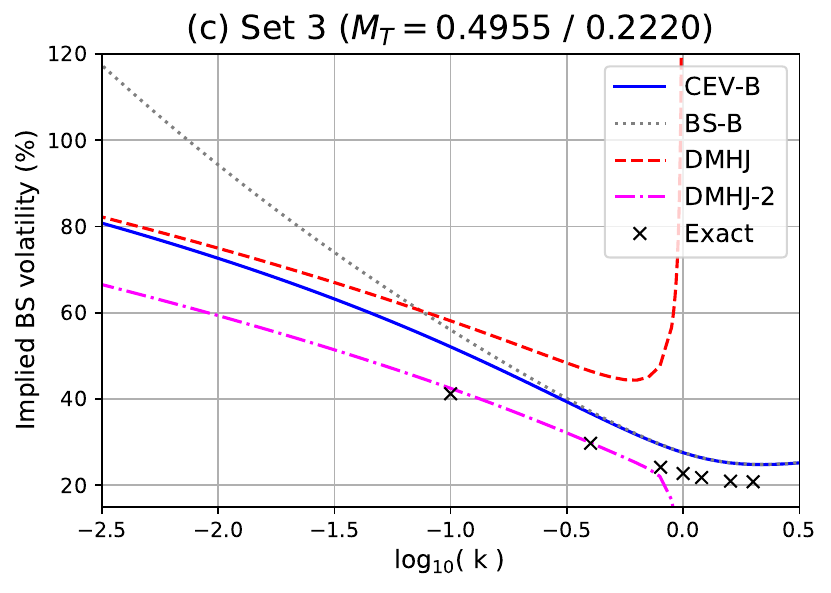}
	\includegraphics[width=0.49\textwidth]{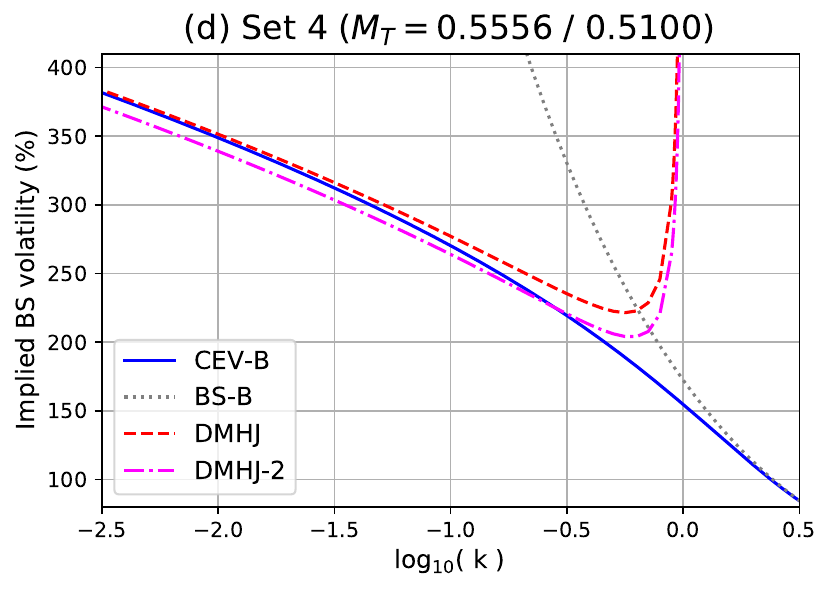}
\end{figure}

Using Sets~1--4, we examine the implication of the mass at zero embedded in the CEV approximations on the volatility smile, and also investigate the arbitrage boundary where negative implied density starts to occur. Figure~\ref{fig:demarco} plots the small-strike BS volatility smile for the four parameter sets. We compare BS-B, CEV-B, and DMHJ along with the exact volatilities from Tables~\ref{tab:set1}--\ref{tab:set3}. The DMHJ asymptotics uses the mass at zero from CEV-B volatility at the origin. In all the sets, BS-B diverges to infinity ($\sim O(|\log k|)$) faster than DMHJ ($\sim O(\sqrt{|\log k|})$) as $k\downarrow 0$, indicating arbitrage. By contrast, CEV-B converges to DMHJ, indicating a lower degree of arbitrage. In near-the-money strike region, CEV-B merges to BS-B instead, while DMHJ diverges at $k=1$. Therefore, CEV-B bridges DMHJ in a low strike and BS-B near the money.

Although the mass at zero implied from Theorem~\ref{th:mass} may not necessarily be accurate, CEV-B is still consistent with DMHJ. To demonstrate this point, we also plot the DMHJ asymptotic smile with a more accurately estimated $M_T$ (DMHJ-2). For Sets~1 and 2, $M_T$ is estimated from MC. For Set~4, $M_T$ is obtained from \citet[Figure~1]{chen2019feeling}. For Set~3, we use an upper bound of $M_T$ obtained from Eq.~\eqref{eq:bound} and the option value at $k=0.1$. We know that this upper bound is tight because the DMHJ-2 smile passes the exact volatilities at $k=0.1$ and 0.4.
Because the mass-at-zero obtained from CEV-B may deviate significantly from the true value, DMHJ and DMHJ-2 can be quite different as shown in Figure~\ref{fig:demarco}(c) for Set~3. Nevertheless, the CEV-B smile is consistent with that of DMHJ up to the possibly incorrect mass at zero. Therefore, we expect CEV-B to exhibit less arbitrage than BS-B. 

To elaborate this point further, we examine the occurrence of arbitrage under the analytic approximations by explicitly detecting the negative probability density function (PDF). The PDF at $k$ is implied from the volatility approximations by the second-order difference equation with small $h$:
$$p(k) = \frac{d^2 C(k)}{d k^2} \approx \frac{C(k+h)-2C(k)+C(k-h)}{h^2},$$
where $C(k)$ is the price of the option struck at $k$, computed using the equivalent volatility at $k$. 
We compare the degree of arbitrage by locating the arbitrage boundary. Given that the analytic approximations exhibit arbitrage at a low strike, the location of the arbitrage boundary is defined as the first $k$ value having a negative implied PDF when $k$ decreases from one to zero. Since the difference equation can also be understood as the premium of the option butterfly, options market makers can trade options above the boundary without being arbitraged against if they use the analytic approximations of the SABR model. Therefore, the lower the arbitrage boundary is, the better the analytic approximation is.

\begin{figure}[ht!]
	\caption{The probability density for \hyperref[tab:param]{Set~3} implied by various approximation methods.} \label{fig:pdf} \vspace{2ex}
	\centering
	\includegraphics[width=0.52\textwidth]{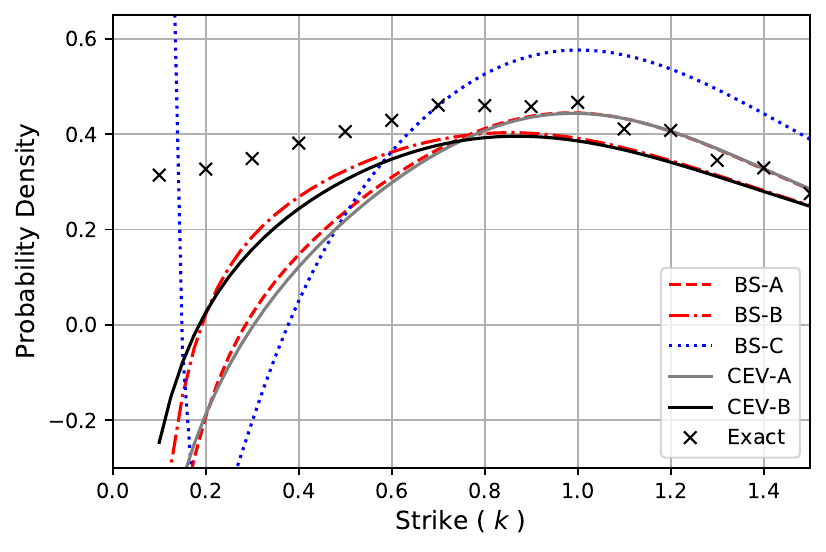}
\end{figure}

Figure~\ref{fig:pdf} depicts the PDF for Set~3 implied from the five methods as well as the exact prices. The PDFs from the approximation methods deviate from the exact value and eventually fall below zero as $k$ approaches zero, indicating arbitrage opportunities at low strikes. The PDFs from BS-B and CEV-B, however, have the lowest arbitrage boundary ($k=0.19$). The PDF deviation of BS-C is the most severe; the PDF falls below zero at the highest strike price ($k=0.30$) and then behaves erratically.

\begin{figure}[ht!]
	\caption{The location of the arbitrage boundary in various approximation methods as a function of $\vov$ (upper panel), $\alpha$ (middle panel), and $\beta$ (lower panel). \hyperref[tab:param]{Set~3} is the base parameter and the base value is indicated by a vertical line. The location of the arbitrage boundary is defined as the first $k$ value at which the implied probability density turns negative as $k$ decreases from one to zero. Therefore, the lower the arbitrage location is, the better the approximation method is.} \label{fig:arb} \vspace{2ex}
	\centering
	\includegraphics[width=0.5\textwidth]{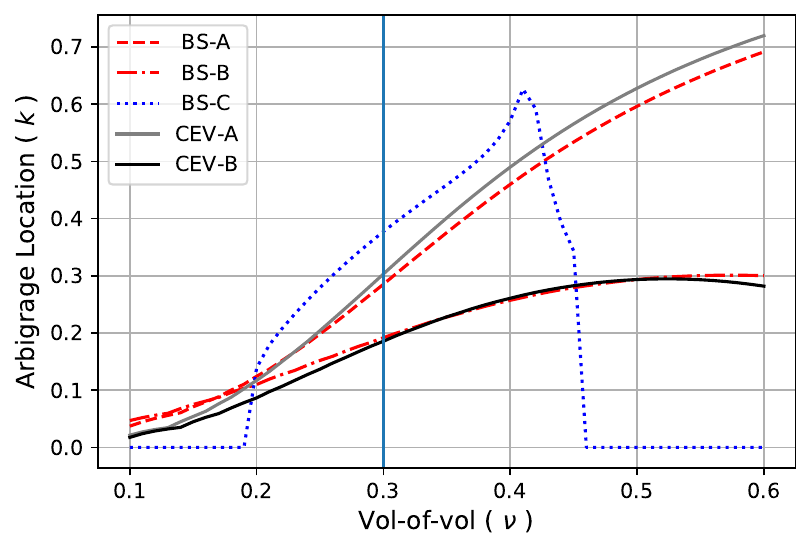}\\
	\includegraphics[width=0.5\textwidth]{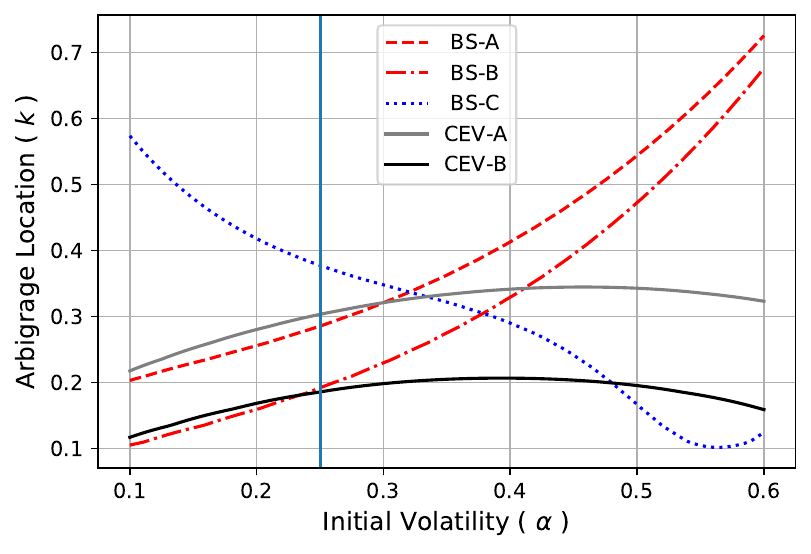}\\
	\includegraphics[width=0.5\textwidth]{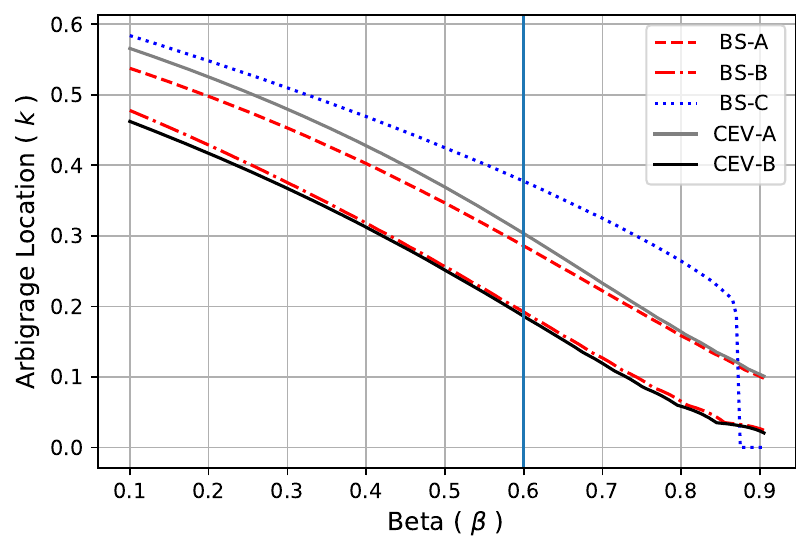}
\end{figure}

We further compare the arbitrage location in wide parameter ranges. In this experiment, we vary $\vov$, $\alpha$, and $\beta$ from the parameter values of Set~3. The upper panel of Figure~\ref{fig:arb} shows the result for varying $\vov$. While the location of the arbitrage boundary generally increases as $\vov$ increases, the arbitrage location is much lower for BS-B and CEV-B than for BS-A and CEV-A. Again, this is because \citet{paulot2015asym}'s refinement becomes pronounced as the $\vov/\alpha$ ratio increases.
In the middle panel where $\alpha$ is varied, the grouping pattern is different; the two CEV-based methods exhibit lower arbitrage boundary than the first two BS-based methods do because the CEV effect dominates \citet{paulot2015asym}'s refinement as $\vov/\alpha$ decreases.
The lower panel shows the result for varying $\beta$. Across all the methods, the arbitrage boundary is higher for $\beta$ closer to zero, reaffirming that the arbitrage is related to the absorbing boundary condition. BS-B and CEV-B have an arbitrage boundary  consistently lower by about 0.1. Overall, CEV-B has the lowest arbitrage boundary among all the methods. CEV-B performs the best in all three parameter variations because the method is equipped with the two enhancements that complement each other: \citet{paulot2015asym}'s refinement and the CEV-based approximation. The arbitrage boundary of BS-C behaves erratically, probably due to the side effect of the higher-order terms in $k$.

\section{Conclusion} \label{sec:conc} \noindent
The SABR model is the dominant stochastic volatility model in the financial industry. Since finding an accurate solution is computationally burdensome, robust analytic approximations of the equivalent volatility are favored in practice despite their imperfection. We show that the quality of the approximation can be significantly improved by deriving the equivalent CEV volatility instead of the BS volatility. Projecting the SABR model on the CEV model takes advantage of an absorbing boundary condition at zero and thus exhibits less arbitrage for small strikes than projecting on the BS model. With this article, we hope to see that the equivalent CEV volatility receives more attention in both research and application.

\section*{Acknowledgements} \noindent
Jaehyuk Choi was supported by the 2019 Bridge Trust Asset Management Research Fund.
Lixin Wu was supported by Grant \#16306717 of the Research Grants Council of Hong Kong. 
The authors thank Nan Chen and Nian Yang for providing the data points of \citet[Figure~1]{chen2019feeling}.

\appendix
\section{Simplification of Paulot's formula} \label{apdx:simp} \noindent
We simplify $t_1$, $t_2$, and $G(t)$ from the original definition of \citet{paulot2015asym}. The intermediate variables in \citet{paulot2015asym} used to define $t_1$, $t_2$, and $G(t)$ are first simplified to
\begin{gather*}
\begin{pmatrix}
x_1 \\ y_1
\end{pmatrix} = \frac{\alpha}{\vov\rhoc}
\begin{pmatrix}
-\rho \\ \rhoc
\end{pmatrix}, \quad
\begin{pmatrix}
x_2 \\ y_2
\end{pmatrix} = \frac{\alpha}{\vov\rhoc}
\begin{pmatrix}
z-\rho\, V \\ \rhoc V
\end{pmatrix}, \\
X = \frac{x_2^2-x_1^2+y_2^2-y_1^2}{2(x_2-x_1)} = \frac{\alpha}{\vov\rhoc}z, \quad 
R = \sqrt{y_1^2+(x_1-X)^2}\, = \frac{\alpha}{\vov\rhoc}V, \\
a = 1, \quad b = \betac\rhoc, \qtext{and} c = \betac\rho.
\end{gather*} 
Using the expressions above, we simplify $t_1$ and $t_2$, respectively, to
\begin{gather*}
t_1 = \sqrt\frac{R-x_1+X}{R+x_1-X} = \sqrt\frac{V+\rho+z}{V-\rho-z} = \frac{V+z+\rho}{\rhoc}, \\
t_2 = \sqrt\frac{R-x_2+X}{R+x_2-X} = \sqrt{\frac{(1+\rho)V}{(1-\rho)V}} = \frac{1+\rho}{\rhoc}.
\end{gather*} 
The function $G(t)$ is also simplified to the form presented in Eq.~\eqref{eq:G}, using
$$ \frac{a+bX}{(1-\beta)R} = \frac{\rhoc\vov k^{\betac}}{\betac\alpha V} = \eta.
$$

\section{Convergence of Paulot's approximation} \label{apx:converge} \noindent
First, we handle the special case of $\beta=1$:
$$ A_2 = \frac{\rho\alpha\vov}{2\rhoc^2}\,\frac{V-1-\rho z}{z^2} = \frac{\rho\alpha\vov}{2\rhoc^2}\, \frac{\rhoc^2}{V+1+\rho z} \rightarrow \frac{\rho}{4}\,\alpha\vov \qtext{as} z\rightarrow 0.
$$
Next, we show that, for $0<\beta<1$, 
$$
\frac{G(t_2) - G(t_1)}{z^2} \rightarrow \frac{\rhoc\betac \alpha}{4 \vov} \qtext{as} z \rightarrow 0.
$$
For all three cases of (\ref{eq:G}), the first and second derivatives are the same as
\begin{gather*}
G'(t) = \frac{1}{1+t^2} - \frac{\eta(\eta - \rhoc)}{\eta^2-1 + (\rho + (\eta - \rhoc)t)^2}, \\
G''(t) = - \frac{2t}{(1+t^2)^2} + \frac{2\eta(\eta-\rhoc)^2(\rho R + (\eta-\rhoc)t)}{[\,\eta^2-1 + (\rho + (\eta-\rhoc)t)^2]^2}.
\end{gather*}
Evaluated at $t=t_2 = (1+\rho)/\rhoc$,
\begin{gather*}
\rho + (\eta-\rhoc)t_2 = \frac{1+\rho}{\rhoc}\eta - 1 \\
(\eta^2-1) + (\rho + (\eta-\rhoc)t_2)^2 = (\eta^2-1) + \left(\frac{1+\rho}{\rhoc}\eta - 1\right)^2
= \frac{2}{1-\rho}\eta(\eta - \rhoc).
\end{gather*}
The first derivative of $G(t)$ at $t=t_2$ is 0, as
$$ G'(t_2) = \frac{1-\rho}{2} - \frac{1-\rho}{2} = 0,
$$
and the second derivative at $t_2$ is 
$$ G''(t_2) = - \frac{\rhoc(1-\rho)}{2} + \frac{(1-\rho)^2}{2\eta}\left(\frac{1+\rho}{\rhoc}\eta - 1\right) = - \frac{(1-\rho)^2}{2\eta} = -\frac{(1-\rho)^2}{2} \frac{\betac \alpha \,V}{\rhoc \vov k^{\betac}}.
$$
We know 
$$ \Delta t = t_2 - t_1 = \frac{1-z-v}{\rhoc} = \frac{-2(1+\rho)z}{\rhoc(1-z+v)} \approx -\frac{1+\rho}{\rhoc}z - \frac{\rhoc}{2}z^2.
$$
Putting all terms together, we have, as $k\rightarrow 1$ ($z\rightarrow 0$),
\begin{gather*}
\frac{G(t_2) - G(t_1)}{z^2} \approx - \frac{G''(t_2) \Delta t^2}{2\,z^2}
= \frac{(1-\rho)^2}{4} \frac{\betac \alpha \,V}{\rhoc \vov k^{\betac}} \frac{(1+\rho)^2}{\rhoc^2}
= \frac{\rhoc\betac \alpha V}{4 \vov k^{\betac}} \rightarrow \frac{\rhoc\betac \alpha}{4 \vov}.
\end{gather*}
Therefore,
$$ A_2 = \frac{\beta\rho\vov^2}{\betac\rhoc}\; \frac{G(t_2)-G(t_1)}{z^2} \rightarrow \frac{\rho\beta}{4} \alpha\vov \qtext{as} k \rightarrow 1. 
$$

\newpage
\singlespacing
\bibliography{../../@Bib/SV_Z2}
\end{document}